\documentclass[jphysa]{iopart}

\usepackage[latin1]{inputenc}
\usepackage{textcomp}
\usepackage{amssymb}
\usepackage{graphicx,xcolor}
\usepackage{amsthm}
\usepackage{hyperref}
\usepackage{braket}
\usepackage{cite}
\usepackage{bm}
\usepackage{relsize}
\usepackage{amsmathred}
\usepackage{multirow}
\usepackage{color}
\usepackage{comment}
\usepackage{times,txfonts}

\def\EGN{$EG_{N} \ $}
\def\M3N{$\mathcal{M}_{N}^{3}$}

\def\EGKM{$EG_{K_M} \ $}
\def\MKA{$\mathcal{M}_{K_\alpha}^{3}$}

\DeclareMathOperator{\conv}{conv}
\newcommand{\odd}{\text{\,odd}}
\newcommand{\even}{\text{\,even}}

\newcommand{\proj}[1]{\ket{#1}\!\bra{#1}}

\mathchardef\ordinarycolon\mathcode`\:
\mathcode`\:=\string"8000
\begingroup \catcode`\:=\active
  \gdef:{\mathrel{\mathop\ordinarycolon}}
\endgroup
\begin{document}

\title{Accessible bounds for general quantum resources}

\author{Thomas R. Bromley, Marco Cianciaruso, Sofoklis Vourekas, Bartosz Regula, and Gerardo Adesso}

\vspace{2mm}

\address{Centre for the Mathematics and Theoretical Physics of Quantum Non-Equilibrium Systems, School of Mathematical Sciences, University of Nottingham, Nottingham NG7 2RD, United~Kingdom}

\vspace{2mm}

\hspace{17mm} {\small E-mail: gerardo.adesso@nottingham.ac.uk}

\begin{abstract}
The recent development of general quantum resource theories has given a sound basis for the quantification of useful quantum effects. Nevertheless, the evaluation of a resource measure can be highly non-trivial, involving an optimisation that is often intractable analytically or intensive numerically. In this paper, we describe a general framework that provides quantitative lower bounds to any resource quantifier that satisfies the essential property of monotonicity under the corresponding set of free operations. Our framework relies on projecting all quantum states onto a restricted subset using a fixed resource non-increasing operation. The resources of the resultant family can then be evaluated using a simplified optimisation, with the result providing lower bounds on the resource contents of any state. This approach also reduces the experimental overhead, requiring only the relevant statistics of the restricted family of states. We illustrate the application of our framework by focusing on the resource of multiqubit entanglement and outline applications to other quantum resources.
\end{abstract}

\section{Introduction}\label{Sec:Int}

Quantum resource theories provide a rigorous structure to characterise the resources present in quantum systems~\cite{horodecki2013quantumness,coecke2016mathematical,brandao2015reversible,gour2017quantum,regula2017convex}. Such resources arise whenever there is a restriction imposed on the available operations that an agent can perform on the quantum system, identifying a set of \emph{free operations} $\mathcal{O}$ which form a subset of the completely positive and trace preserving linear maps~\cite{nielsen2010quantum}. The restriction also identifies a set of \emph{free states} $\mathcal{F}$, forming the largest subset of the set $\mathcal{D(H)}$ of quantum states for which any pair of states can be reversibly converted using free operations alone. Any non-free state is hence a \emph{resource state}, since one must always  input resource in the form of a non-free operation to create such a state from a free state.%, while any non-free operation is a \emph{resource consuming operation}.

The restricted agent based approach~\cite{del2015resource} to characterising quantum resources has already been particularly fruitful for understanding quantum entanglement~\cite{horodecki2009quantum,bengtsson2007geometry}, where the restriction is given by the paradigm of local operations and classical communication (LOCC) between spatially separated parties~\cite{bennett1996concentrating,horodecki2009quantum,donald2002uniqueness}.
In fact, entanglement theory can act as a progenitor for modelling more general resource theories. For example, the many-copy interconversion between resource states using free operations, first understood for entanglement theory, leads to the general concept of resource distillation and cost~\cite{bennett1996concentrating,bennett1996mixed,rains1999bound,rains1999rigorous,brandao2013resource,anshu_2017}. The development of general quantum resource theories has led to further understanding of the resources of quantum coherence~\cite{baumgratz2014quantifying,streltsov2017colloquium,aberg2006quantifying}, quantum correlations~\cite{adesso2016measures,bromley2016there,modi2012classical,streltsov2011behavior,hu2012necessary,guo2013necessary,brodutch2012criteria,aaronson2013comparative,spehner2014quantum} and other nonclassical properties~\cite{brandao2015second,gour2015resource,veitch2014resource,schuch2004nonlocal,gour2008resource,de2014nonlocality,liu2016theory,devetak2008resource,theurer2017resource,ahmadi2017quantification}.

An important question is to consider how much resource is present in a given state. The free operations allow for a qualitative characterisation: a state $\rho$ is more resourceful than another state $\sigma$ if there exists a free operation $\Lambda \in \mathcal{O}$ so that $\sigma = \Lambda(\rho)$, meaning that the state $\sigma$ can be prepared from $\rho$ without consuming any resource. Such a characterisation can result in a complicated multibranch hierarchy~\cite{dur2000three,verstraete2002four}, where it can be difficult to identify necessary and sufficient conditions for interconvertibility between two resource states~\cite{nielsen2010quantum,nielsen1999conditions,hayden2000locc,chitambar2016critical,du2015conditions,winter2016operational}.

%has been accentuated by the promise of the evolution of quantum technologies of the coming years~\cite{dowling2003quantum,o2009photonic,ursin2007entanglement,bouwmeester1997experimental,brida2010experimental,ladd2010quantum}, which necessarily harness purely quantum properties to achieve a supremacy over presently available technologies~\cite{preskill2012quantum}.

However, quantum resource theories also provide the structure to quantitatively measure the resource content of a state~\cite{plenio2007introduction,bengtsson2007geometry,adesso2016measures,streltsov2017colloquium,regula2017convex}. Here, the complicated multibranch hierarchy can be condensed into a single quantitative ordering that preserves the hierarchy within each branch. Since there is not a unique way to impose a quantitative ordering, there is no unique measure of the resources present in a quantum system. Although this may appear counterintuitive, one may reconcile the non-uniqueness of resource measures from an operational perspective: we expect to exploit our quantum resource for a variety of different tasks, for which each task may value certain resource states over others and hence impose a different ordering.

Any \emph{bona fide} measure ${R}$ of a quantum resource must have compatibility with the corresponding quantum resource theory by satisfying two universal requirements. First, it must hold that ${R}(\rho)\geq 0$ for all $\rho \in \mathcal{D(H)}$ and ${R}(\rho)= 0$ for all $\rho \in \mathcal{F}$, i.e. that a resource measure is in general non-negative and always zero when there is no resource. Second, it must hold that ${R}(\Lambda(\rho))\leq {R}(\rho)$ for all $\rho \in \mathcal{D(H)}$ and $\Lambda \in \mathcal{O}$. This requirement is known as \emph{resource monotonicity}, and imposes that resource measures should preserve the hierarchy within each branch. Additional properties may also be considered for a given resource, such as strong monotonicity~\cite{horodecki2009quantum} or convexity whenever $\mathcal{F}$ is convex (see e.g.~\cite{bengtsson2007geometry,plenio2007introduction} for comprehensive accounts of the requirements for measures of entanglement).

When a bona fide resource measure is selected, one is then presented with the task of evaluating the measure for arbitrary states. This task is typically intractable analytically and difficult numerically, often resulting in strong restrictions on the applicability of the resource measure. For example, consider the non-trivial optimisations given by the distance-based resource measures
\begin{equation}\label{Eq:DistBased}
{R}^{D_{\delta}}(\rho) := \inf_{\sigma \in \mathcal{F}} D_{\delta}(\rho,\sigma),
\end{equation}
where $D_{\delta}$ is a contractive distance on the set of quantum states~\cite{bengtsson2007geometry,vedral1997quantifying,vedral1998entanglement}, as well as by the (generalised) resource robustness~\cite{vidal1999robustness,steiner2003generalized,napoli2016robustness,piani2016robustness,bromley2017navigating,regula2017convex}
\begin{equation}\label{Eq:RobustnessGeneral}
{R}^{R}(\rho):= \inf_{\tau \in \mathcal{D(H)}} \left\lbrace s \geq 0 \left| \frac{\rho + s \tau}{1+s}=: \sigma \in \mathcal{F} \right\rbrace \right. ,
\end{equation}
which quantifies the resilience of a resource state $\rho$ against mixing.

In this paper, we construct a general  framework to calculate simplified lower bounds to bona fide resource measures. %Our framework may be applied to any resource measure satisfying the resource monotonicity requirement.
We begin in Section~\ref{Section:RNIPS} by introducing two of the foundational concepts for our framework: the resource non-increasing projections and the corresponding resource guarantor states, both of which can have wider relevance to quantum resource theory. In Section~\ref{Section:Framework}, we detail the four main steps of our framework. Our approach is not restricted to specific types of resource, since it relies on general concepts using the structure of resource theories. Nevertheless, to verify the usability of our framework, we provide example applications in Section~\ref{Section:Application}, focusing in particular on the resource of multiqubit entanglement. Here we first provide a method for constructing entanglement non-increasing projections and identifying their corresponding entanglement guarantor states. By using this construction we define a new family of entanglement guarantor states complementing those highlighed in~\cite{cianciaruso2016accessible}, and proceed to evaluate the robustness of multiqubit entanglement on these states, which can in turn be used to lower bound the robustness of the GHZ state. Finally, we conclude and discuss our findings in Section~\ref{Section:Discussion}.%, including extensions of our work by identifying other resource non-increasing projections.

\section{Resource non-increasing projections and resource guarantor states}\label{Section:RNIPS}
\begin{figure}[t]
    \centering
    \includegraphics[width=0.6\textwidth]{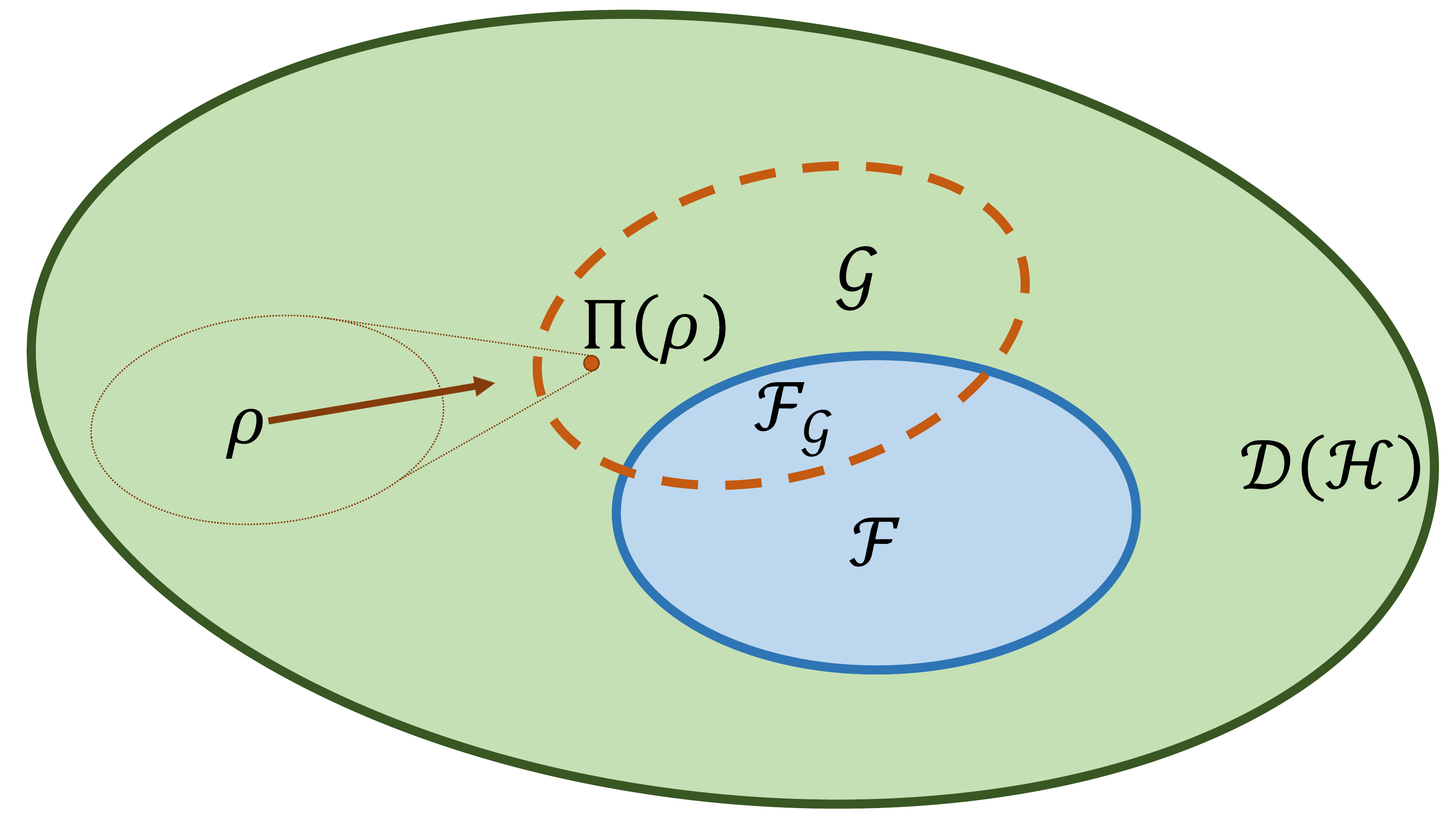}
    \caption{The action of a resource non-increasing projection (RNIP) $\Pi$, which satisfies $\Pi^2 = \Pi \in \mathcal{O}$, is to project the set of states $\mathcal{D(H)}$ onto the set of resource guarantor states (RGSs) $\mathcal{G}$ (dashed orange ellipse). Any state $\rho \in \mathcal{D(H)}$ has a corresponding RGS $\Pi(\rho)\in \mathcal{G}$ (orange circle), with a many-to-one correspondence between general states and RGSs (dotted orange area). The intersection between free states $\mathcal{F}$ (solid blue ellipse) and RGSs $\mathcal{G}$ is the set of free RGSs $\mathcal{F}_{\mathcal{G}}$, so that every free state $\sigma \in \mathcal{F}$ is transformed into a corresponding free RGS $\Pi(\sigma) \in \mathcal{F}_{\mathcal{G}}$.}
    \label{Figure:RNIPsAndRGSs}
\end{figure}

We now introduce the two main foundations of our framework. A quantum operation $\Pi$ that satisfies the composition relation $\Pi^2 = \Pi$ is referred to as a \emph{projection}\footnote{Here we restrict to quantum operations that preserve the dimension of the quantum system.}. We define the \emph{resource non-increasing projections} (RNIPs) to be the subset of projections that are also free. Every such $\Pi \in \mathcal{O}$  identifies a corresponding set of \emph{resource guarantor states} (RGSs) $\mathcal{G}$ given by all the states left invariant by $\Pi$, i.e.
\begin{equation}
\mathcal{G} = \left\lbrace \rho \in \mathcal{D(H)} \,\,\, \left| \,\,\, \Pi(\rho) = \rho  \right\rbrace \right. .
\end{equation}
It can then be seen that the action of $\Pi$ on the set of quantum states $\mathcal{D(H)}$ is to project every state onto the set of resource guarantor states, so that
\begin{align}
\begin{split}
\Pi(\rho) &\in  \mathcal{G} \qquad \forall \,\,\, \rho \in \mathcal{D(H)},  \\
\Pi(\rho) &= \rho \qquad \, \forall \,\,\, \rho \in \mathcal{G}.
\end{split}
\end{align}

Hence, for every $\rho \in \mathcal{D(H)}$ there is a corresponding RGS $\Pi(\rho) \in \mathcal{G}$. Now, for any bona fide measure of a resource ${R}$ we know that
\begin{equation}\label{Eq:RGSLowerBound}
{R}(\Pi (\rho)) \leq {R}(\rho),
\end{equation}
which holds since ${R}$ satisfies the resource monotonicity requirement. It can therefore be seen that the state $\Pi(\rho) \in \mathcal{G}$ provides a quantitative guarantee on the resources of $\rho \in \mathcal{D(H)}$ in terms of a lower bound. Figure~\ref{Figure:RNIPsAndRGSs} illustrates the action of RNIPs and the corresponding set of RGSs. We remark that there is generally a many-to-one correspondence between a general state $\rho \in \mathcal{D(H)}$ and the corresponding RGS $\Pi(\rho) \in \mathcal{G}$. Furthermore, RNIPs present a generalisation of the resource destroying maps introduced in~\cite{liu2016theory}, which are an extremal form of RNIPs that destroy all resource, as their  RGSs are free states.

In general, since $\Pi \in \mathcal{O}$, we know that the action of $\Pi$ on the set of free states $\sigma \in \mathcal{F}$ is to map it to a subset $\mathcal{F}_{\mathcal{G}}$ of free RGSs, i.e. the intersection between $\mathcal{F}$ and $\mathcal{G}$. We have that
\begin{equation}\label{Eq:FreeRNGs}
\mathcal{F}_{\mathcal{G}} = \left\lbrace \Pi(\rho) \,\,\, \left| \,\,\, \rho \in \mathcal{F}  \right\rbrace \right. \subset \mathcal{F}.
\end{equation}
We see in the following that our framework allows for a simplification in the evaluation of resource measures by replacing optimisation over all free states $\mathcal{F}$ with an optimisation over the free RGSs $\mathcal{F}_{\mathcal{G}} \subset \mathcal{F}$, which are typically simpler to characterise.

\section{Framework}\label{Section:Framework}

\begin{figure}[t]
    \centering
    \includegraphics[width=0.6\textwidth]{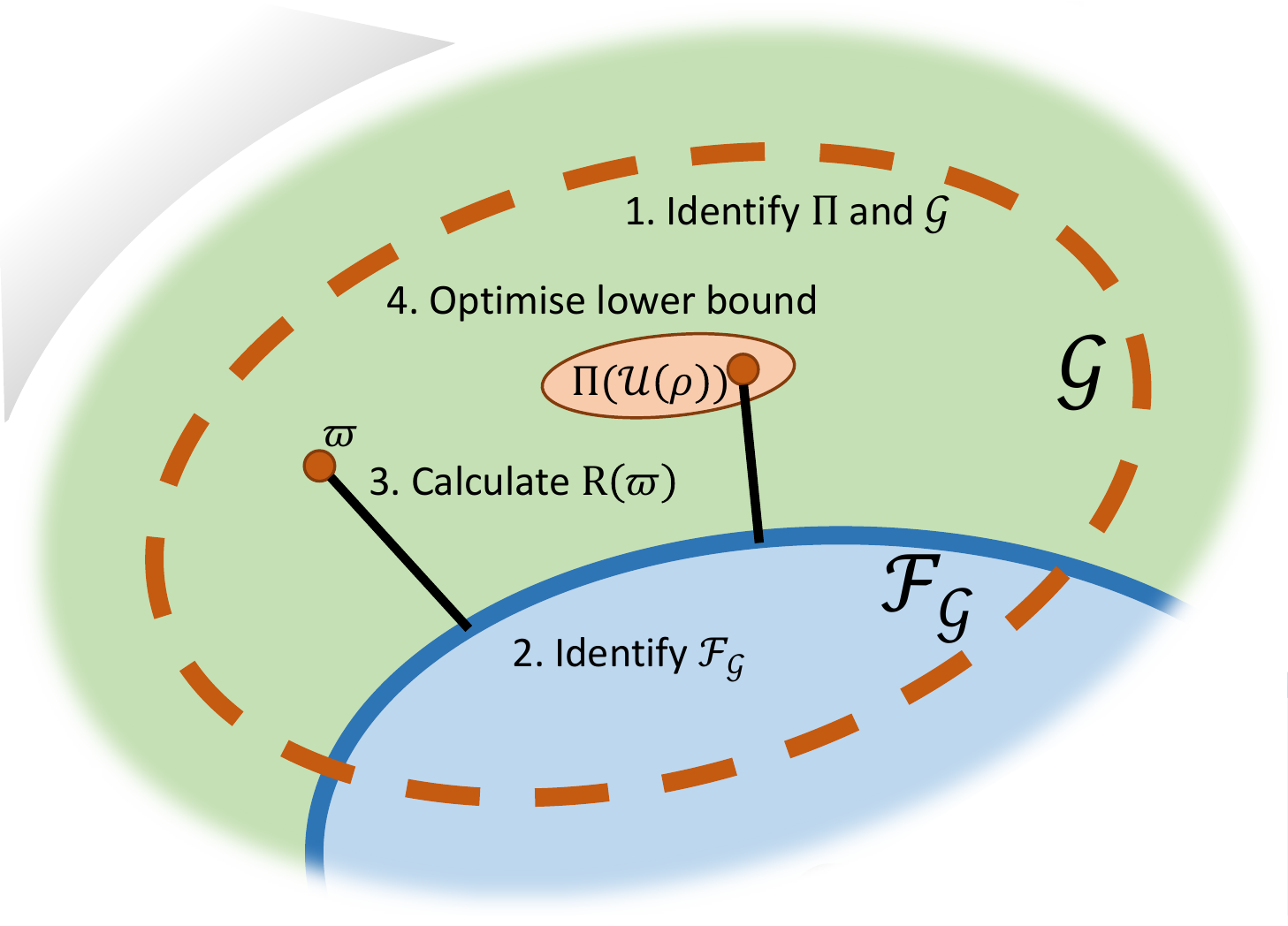}
    \caption{A zoom-in onto the set $\mathcal{G}$ of RGSs illustrates the four steps of our framework. The first two steps consist of fixing an RNIP $\Pi$ and finding the sets of RNGs $\mathcal{G}$ (dashed orange ellipse) and free RNGs $\mathcal{F}_{\mathcal{G}}$ (given by the intersection with the solid blue ellipse). In the third step, we use the characterisations of $\mathcal{G}$ and $\mathcal{F}_{\mathcal{G}}$ to evaluate a chosen resource measure ${R}(\varpi)$ for any RGS $\varpi \in \mathcal{G}$ (illustrated here for the distance-based measures in Eq.~(\ref{Eq:DistanceBasedRGS})). Finally, the fourth step involves considering the optimised lower bound $\max_{\mathcal{U}} {R}(\Pi(\mathcal{U}(\rho)))\leq {R}(\rho)$ over all free unitary operations $\mathcal{U} \in \mathcal{O}$, with the set $\Pi(\mathcal{U}(\rho)) \in \mathcal{G}$ of RGSs illustrated by the orange ellipse.}
    \label{Figure:Framework}
\end{figure}

We now specify the general framework for providing lower bounds to bona fide resource measures for arbitrary quantum states. Our framework consists of four steps.
\begin{description}
\item[Step One: ]{Identify an RNIP $\Pi$ and characterise the corresponding set of RGSs $\mathcal{G}$.}
\item[Step Two: ]{Characterise the set $\mathcal{F}_{\mathcal{G}}$ of free RGSs.}
\item[Step Three: ]{Evaluate ${R}(\varpi)$ for all $\varpi \in \mathcal{G}$.}
\item[Step Four: ]{Optimise the lower bound ${R}(\Pi(\mathcal{U}(\rho))) \leq {R}(\rho)$ over free unitaries $\mathcal{U} \in \mathcal{O}$.}
\end{description}
Each step is now explained in detail. An illustration of the framework is provided in Fig.~\ref{Figure:Framework} and an example of its application to the resource of multiqubit entanglement can be found in Sec.~\ref{Section:Application}.

The first step is to identify an RNIP and characterise the corresponding RGSs. %We give a specific example of how this may be achieved for the resource of multiqubit entanglement in Sec.~\ref{Section:Application}.
This step requires attention to two objectives: on the one hand it is desirable for the RGSs and free RGSs to be simple to characterise, so that ${R}(\varpi)$ can be evaluated for any $\varpi \in \mathcal{G}$. On the other hand, one does not want to pick an RNIP that destroys a lot of resource, as the resultant lower bound in Eq.~(\ref{Eq:RGSLowerBound}) becomes less informative. Indeed, it is possible for $\Pi(\rho) \in \mathcal{F}_{\mathcal{G}}$ even if $\rho \notin \mathcal{F}$, so that the corresponding lower bound is trivial. This is seen in the extreme for the resource destroying maps~\cite{liu2016theory}, for which $\Pi(\rho) \in \mathcal{F}_{\mathcal{G}}$ for all $\rho \in \mathcal{D(H)}$.

The second step consists of characterising the set $\mathcal{F}_{\mathcal{G}}$ of free RGSs, i.e. the intersection between $\mathcal{F}$ and $\mathcal{G}$. This can be achieved using Eq.~(\ref{Eq:FreeRNGs}), which tells us that $\mathcal{F}_{\mathcal{G}}$ is simply the result of applying the chosen RNIP onto the set of  free states $\mathcal{F}$.

In the third step of the framework, we evaluate a chosen resource measure ${R}(\varpi)$ on the set of RGSs, i.e. for all $\varpi \in \mathcal{G}$. Typically, the evaluation of ${R}(\varpi)$ for $\varpi \in \mathcal{G}$ is much more affordable than the evaluation of ${R}(\rho)$ for $\rho \in \mathcal{D(H)}$, since one can employ a number of tricks to simplify the optimisation. For example, consider the distance-based resource measures in Eq.~(\ref{Eq:DistBased}). It can be seen that for any $\varpi \in \mathcal{G}$
\begin{eqnarray}\label{Eq:DistanceBasedRGS}
{R}^{D_{\delta}}(\varpi) &=& \inf_{\sigma \in \mathcal{F}} D_{\delta}(\varpi,\sigma) \nonumber \\
&=& \inf_{\sigma \in \mathcal{F}} D_{\delta}(\Pi(\varpi),\Pi(\sigma)) \nonumber \\
&=& \inf_{\sigma \in \mathcal{F}_{\mathcal{G}}} D_{\delta}(\varpi,\sigma),
\end{eqnarray}
where in the second equality we use the contractivity of the distance $D_{\delta}$ under any quantum operation, while in the third equality use Eq.~(\ref{Eq:FreeRNGs}) and the fact that $\Pi(\varpi) = \varpi$. This equation means that the distance-based resources of $\varpi \in \mathcal{G}$ are given simply by the distance to the free RGSs.

Alternatively, consider the resource robustness in Eq.~(\ref{Eq:RobustnessGeneral}). Whenever we consider an RGS $\varpi \in \mathcal{G}$, for every mixture $\frac{\varpi + s \tau}{1+s}\in \mathcal{F}$ with $\tau \in \mathcal{D(H)}$, there is a corresponding  mixture $\frac{\varpi + s \Pi(\tau)}{1+s}\in \mathcal{F}_{\mathcal{G}}$. Hence, it holds that
\begin{equation}\label{Eq:RobustnessRGS}
{R}^{R}(\varpi) = \inf_{\tau \in \mathcal{G}} \left\lbrace s \geq 0 \left| \frac{\varpi + s \tau}{1+s}=: \sigma \in \mathcal{F}_{\mathcal{G}} \right\rbrace \right. ,
\end{equation}
so that one needs only to consider mixtures of $\varpi$ with other RGSs to obtain a free RGS. We note that Eq.~(\ref{Eq:RobustnessRGS}) is a convex optimisation problem whenever $\mathcal{G}$ and $\mathcal{F}_{\mathcal{G}}$ are convex sets, and that Eq.~(\ref{Eq:RobustnessRGS}) may be evaluated as the solution of a semidefinite program~\cite{boyd2004convex,piani2016robustness}  if $\mathcal{G}$ and $\mathcal{F}_{\mathcal{G}}$ can additionally be characterised with a finite number of linear matrix inequalities. This can be the case even when Eq.~(\ref{Eq:RobustnessGeneral}) cannot be posed as the solution to a semidefinite program, as we see for the example in Sec.~\ref{Section:Application} of multiqubit entanglement.

The final step of our framework is to provide a lower bound to the resource degree of any state $\rho \in \mathcal{D(H)}$, according to Eq.~(\ref{Eq:RGSLowerBound}), by considering the corresponding RGS $\Pi(\rho) \in \mathcal{G}$. However, this lower bound may be optimised over free unitaries: the unitary transformations $\mathcal{U}(\rho):= U \rho U^{\dagger}$ with $U U^{\dagger} = U^{\dagger} U = \mathbb{I}$ such that $\mathcal{U} \in \mathcal{O}$. Indeed, it is straightforward to see that ${R}(\rho) = {R}(\mathcal{U(\rho)})$ for any monotonic resource measure, so that
\begin{equation}
{R}(\Pi(\mathcal{U}(\rho))) \leq {R}(\mathcal{U}(\rho)) = {R}(\rho).
\end{equation}
Evaluating the maximum of the left hand side of the above equation over all free unitary operations $\mathcal{U} \in \mathcal{O}$ hence provides an optimised lower bound to the resource contents of $\rho \in \mathcal{D(H)}$. Nevertheless, the set of free unitaries may not always be fully characterised for a given resource, while optimisation over the free unitaries can be computationally intensive. It can thus be more realistic to optimise over a well characterised subset of free unitaries. As we see in the following, the free unitaries of multiqubit entanglement are local qubit unitaries. Performing an optimisation over the ${\sf SU(2)}$ group is often simple to solve, with some cases amenable to evaluation as the result of a semidefinite program.%~\cite{} (can clear this part up for robustness of M3N, it is quite a nice question).

These four steps compose the main structure of our framework. Whilst step one and step four have already been employed for the resource of entanglement~\cite{buchholz2015evaluating,hofmann2014analytical,siewert2012quantifying,eltschka2012quantitative}, our primary contribution here is formalising the framework for general resources, as well as highlighting the simplifications possible in evaluating resource measures by restricting to what we have defined as RGSs, see Eqs.~(\ref{Eq:DistanceBasedRGS}) and~(\ref{Eq:RobustnessRGS}).

It is also important to comment on the experimental applicability of our result.  One approach to quantify the resource of a state prepared in the laboratory is to perform a full state tomography~\cite{james2001measurement,thew2002qudit,nielsen2010quantum}, requiring an exponential number of measurement settings (with respect to system dimension) in the worst-case scenario, although a less intensive overhead can be achieved by restricting to low rank states\cite{gross2010quantum}. Alternatively, one can reconstruct the corresponding RGS, which may require much fewer measurement settings to achieve. For example, in the following section we discuss a family of RGSs that are experimentally accessible using only three local measurement settings. Moreover, the optimisation over free unitaries $\mathcal{U} \in \mathcal{O}$ in the fourth step of our framework can be attempted experimentally whenever partial prior knowledge of the target state is available, as is the case in most scenarios. Here, one performs a change of basis before the measurement according to the optimal free unitary of the target state.

\section{Applications of the framework}\label{Section:Application}

The framework naturally lends itself to the characterisation of a variety of quantum resources. We first briefly discuss some very natural example applications for thermodynamics and quantum coherence, before proceeding to give a detailed example of applying our framework to multiqubit entanglement.

Our first example, the resource theory of thermodynamics (or athermality) \cite{brandao2013resource,gour2015resource}, identifies a unique free state --- the thermal state of a given Hamiltonian $H$ at a fixed temperature --- and the free operations as those which can be implemented by attaching a thermal ancilla and applying a unitary operation which commutes with the total Hamiltonian of the system. Here, one can consider as a resource non-increasing projection the completely dephasing map $\Delta_H(\cdot) = \sum_i \proj{E_i} \cdot \proj{E_i}$ where $\{\ket{E_i}\}$ is the eigenbasis of $H$. This greatly simplifies the evaluation of resource measures, since the resultant states are simply classical probability distributions, and indeed the problem of computing distance-based resource measures reduces to optimising distances between probability distributions. Such projections have already found use in the description of operational tasks in this resource theory \cite{brandao2013resource,horodecki_2013-1,lostaglio_2015}.

Another example is quantum coherence, which captures the existence of a quantum system in a superposition of states with respect to a given reference basis, and has relevance in fundamental information processing tasks, metrology, and quantum biology, as well as being a crucial ingredient for the creation of entanglement~\cite{streltsov2017colloquium}. The coherence-free states, or \emph{incoherent states}, can be identified as diagonal when represented in the reference basis $\{\ket{i}\}$, while the free operations can be identified as \emph{incoherent operations}~\cite{baumgratz2014quantifying}. Any projective measurement on subspaces spanned by vectors of the reference basis, representing a decohering map which zeroes some off-diagonal elements of the density matrix, is then a resource non-increasing projection. Identifying particular instances of this type of projection will then provide varying lower bounds for coherent states. Another type of an RNIP which has been employed in the resource theory of coherence is an operation reducing all two-qubit states to so-called \M3N states~\cite{silva2016observation}, which we will also encounter in Sec. \ref{sec:applying_entanglement}.

%However, the structure of the resources considered above is relatively simple. To exemplify the usefulness of our framework in significantly more complex applications,
We now provide a more in-depth analysis of the application of our framework to the resource theory of multiqubit entanglement --- a fundamental resource of paramount importance in quantum information \cite{horodecki2009quantum}, although quantitatively very difficult to characterise. We begin by outlining the background details of multiqubit entanglement and then proceed to discuss a general method to construct entanglement non-increasing projections and find their corresponding entanglement guarantor states. This construction is used to identify the \EGN states, which we then use within our framework.

\subsection{Resource theory of multiqubit entanglement}

Within the quantum resource theory of multiqubit entanglement, there exists a hierarchy of free states referred to as $M$\textit{-separable states}, with $2 \leq M \leq N$. These states can be written as convex combinations of product states, each of which is factorised with respect to any (possibly different) partition of the $N$ qubits into $M$ subsystems, i.e.
\begin{equation}\label{Eq:MSep}
\varsigma = \sum_i p_i |\psi_i^{(1)}\rangle\langle\psi_i^{(1)}|\otimes|\psi_i^{(2)}\rangle\langle\psi_i^{(2)}|\otimes\cdots\otimes|\psi_i^{(M)}\rangle\langle\psi_i^{(M)}|,
\end{equation}
where $|\psi_i^{(\alpha)}\rangle$ is any pure state of the $\alpha$-th subsystem of the $M$-partition corresponding to the $i$-th term (we stress that the $M$-partition is allowed to vary for different values of $i$). The hierarchy of $M$-separable states stands as follows: $M$-separability implies $M'$-separability for any $M'<M$, whereas $M$-inseparability implies $M'$-inseparability for any $M'>M$. For example, when considering the two extremes of this hierarchy, we have that $N$-separability implies any other form of $M$-separability, and is thus called \textit{full separability}, whereas $2$-inseparability implies any other form of $M$-inseparability, and is thus called \textit{genuine multiqubit entanglement} or \textit{full inseparability}.

The free operations are instead given by the single-qubit LOCC, whereby only operations that are local on each of the $N$ qubits  are permitted, along with classical communication~\cite{chitambar2014everything}. An important instance of single-qubit LOCC is a convex combination of single-qubit local unitaries, whose action on a state $\rho$ is given by
\begin{equation}
\sum_{i} p_{i}\  U_{i}^{(1)} \otimes U_{i}^{(2)} \otimes \ldots \otimes U_{i}^{(N)} \rho\  U_{i}^{(1) \dagger} \otimes U_{i}^{(2) \dagger} \otimes \ldots \otimes U_{i}^{(N) \dagger}.
\end{equation}
It requires only one-way communication and can be physically achieved by allowing one of the qubit parties, e.g. the $\alpha$-th one, to randomly select a local unitary $U_{i}^{(\alpha)}$ by using the probability distribution $\{p_{i}\}$ and then to communicate the result to all the other parties.

Having identified the free states and free operations, we can define a bona fide measure $E_{M}$ of $M$-inseparable multiqubit entanglement to be any function satisfying the requirements discussed in Sec.~\ref{Sec:Int}. In particular, the distance-based measures $E_{M}^{D_{\delta}}$ are specified by Eq.~(\ref{Eq:DistBased}) and the entanglement robustness is specified by Eq.~(\ref{Eq:RobustnessGeneral}).

\subsection{Constructing entanglement non-increasing projections}\label{Sec:Systematic}

We now introduce a systematic way to build multiqubit entanglement non-increasing projections (ENIPs) and, as a consequence, the corresponding entanglement guarantor states (EGSs). First of all, let us give a shorthand notation for the Bloch representation of a generic $N$-qubit state $\rho$ in the computational basis $\{|0\rangle,|1\rangle \}^{\otimes N}$:
\begin{equation}\label{eq:BlochRepresentation}
\rho=\frac{1}{2^N} \sum_{\bm{\alpha}\in I_N} T^\rho_{\bm{\alpha}} P_{\bm{\alpha}},
\end{equation}
where the set $I_N=\{0,1,2,3\}^{N}$ contains all the $N$-tuples $\bm{\alpha}=(\alpha_1,\alpha_2,\cdots,\alpha_N)$ of indices ranging from $0$ to $3$, $P_{\bm{\alpha}}=\sigma_{\alpha_1}\otimes\sigma_{\alpha_2}\otimes\cdots\otimes\sigma_{\alpha_N}$, with $\sigma_0=\mathbb{I}$ being the $2\times 2$ identity matrix and $\sigma_1,\sigma_2,\sigma_3$ the Pauli matrices, and $T^\rho_{\bm{\alpha}}=\mbox{Tr}(\rho P_{\bm{\alpha}})$ are the so-called correlation tensor elements of $\rho$. The single-qubit (Hermitian) local unitaries $P_{\bm{\alpha}}$ satisfy several properties that will be extremely useful in the following, see \ref{Appendix:Palpha} for further details.

Now we provide a systematic way to project, via single-qubit LOCC, an arbitrary state $\rho$ onto an EGS of the following form:
\begin{equation}\label{eq:entanglementguarantorstates}
\varpi_\rho = \frac{1}{2^N} \sum_{\bm{\alpha}\in G} T^\rho_{\bm{\alpha}} P_{\bm{\alpha}},
\end{equation}
for some instances of $G\subset I_N$. This ENIP consists of putting equal to zero all the correlation tensor elements $T^\rho_{\bm{\alpha}}$ of $\rho$ such that $\bm{\alpha}\notin G$ and leaving alone the remaining ones. The number of surviving correlation tensor elements $T^\rho_{\bm{\alpha}}$ is given by the cardinality $|G|$ of the set $G$. One may pick $G$ so that $|G|$ is small, leading to a reduction in the complexity of evaluating the multiqubit entanglement of $\varpi_{\rho}$ as well as a decreased overhead in the number of local measurement settings required to recover $\varpi_{\rho}$ in laboratory. However, $|G|$ can be large and still tractable, e.g., for the entanglement robustness whenever $\varpi_{\rho}$ and the corresponding free states can be described with a finite number of linear matrix inequalities.

One approach to performing the above ENIP is to apply to $\rho$ the following convex combination of single-qubit local unitaries (which is a single-qubit LOCC as we have previously mentioned):
\begin{equation}\label{eq:groupavarageprojection}
\Pi_{G}(\rho):= \frac{1}{|J_G|} \sum_{\bm{\alpha}\in J_G} P_{\bm{\alpha}}' \rho P_{\bm{\alpha}}'^{\dagger},
\end{equation}
where $J_G\subset I_N$ is defined in such a way that
\begin{equation}\label{eq:conditionsonUG}
\left\{
\begin{array}{c}
[P_{\bm{\alpha}},P_{\bm{\beta}}']=0 \ \ \ \ \forall \bm{\alpha}\in G, \ \bm{\beta}\in J_G, \\
\exists \bm{\beta}\in J_G : \{P_{\bm{\alpha}},P_{\bm{\beta}}'\}=0 \ \forall \bm{\alpha} \notin G .
\end{array}
\right.
\end{equation}
This ENIP is successful, i.e. so that $\Pi_{G}(\rho)$ can always be written as in Eq.~(\ref{eq:entanglementguarantorstates}), provided that the matrices $P_{\bm{\beta}}'$ for which $\bm{\beta}\in J_G$ form a set that can be written as
\begin{equation}\label{eq:groupuptoaphase}
\{P_{\bm{\beta}_i}'\}_{i=1}^{2^{n}} =
\left \{
  \begin{tabular}{c}
  $\mathbb{I}^{\otimes N}$ \\
  $\{P_{\bm{\alpha}_{i_1}}\}_{i_{1}=1}^{n}$ \\
  $\{P_{\bm{\alpha}_{i_2}}P_{\bm{\alpha}_{i_1}}\}_{i_{2}>i_{1}=1}^{n}$ \\
  $\cdots$ \\
  $\{P_{\bm{\alpha}_{i_n}} \ldots P_{\bm{\alpha}_{i_2}}P_{\bm{\alpha}_{i_1}}\}_{i_{n}>\ldots>i_{2}>i_{1}=1}^{n}$ \\
  \end{tabular}
\right \},
\end{equation}
for some family of matrices $\{P_{\bm{\alpha_i}}\}_{i=1}^{n}$.%NOT TRUE (note that $|J_G| = 2^{n}$).

Indeed, by using the properties discussed in \ref{Appendix:Palpha}, one can easily see that
\begin{equation}
\Pi_{G}(\rho) = \varpi_\rho,
\end{equation}
where $\varpi_\rho$ is defined in Eq.~(\ref{eq:entanglementguarantorstates}). This is due to the fact that for any $\bm{\alpha} \in I_{N}$
\begin{eqnarray}
\mbox{Tr}(\Pi_{G}(\rho)P_{\bm{\alpha}}) &=& \frac{1}{|J_G|} \sum_{\bm{\beta}\in J_G} \mbox{Tr} (P_{\bm{\beta}}' \rho P_{\bm{\beta}}'^{\dagger} P_{\bm{\alpha}}) \nonumber \\
&=&  \frac{1}{|J_G|} \sum_{\bm{\beta}\in J_G} \mbox{Tr} (\rho P_{\bm{\beta}}'^{\dagger} P_{\bm{\alpha}} P_{\bm{\beta}}') \nonumber \\
&=&  \frac{1}{|J_G|} \left[ \sum_{\bm{\beta}\in J_G^+(\bm{\alpha})} \mbox{Tr} (\rho P_{\bm{\alpha}} P_{\bm{\beta}}'^{\dagger}P_{\bm{\beta}}')  - \sum_{\bm{\beta}\in J_G^-(\bm{\alpha})}\mbox{Tr} (\rho P_{\bm{\alpha}} P_{\bm{\beta}}'^{\dagger}P_{\bm{\beta}}')\right] \nonumber \\
&=& \frac{1}{|J_G|} (|J_G^+(\bm{\alpha})|-|J_G^-(\bm{\alpha})|) \mbox{Tr} (\rho P_{\bm{\alpha}}) \nonumber \\
&=& \left(2 \frac{|J_G^+(\bm{\alpha})|}{|J_G|} - 1\right) \mbox{Tr} (\rho P_{\bm{\alpha}})  \nonumber \\
&=& \left\{
\begin{array}{c}
\mbox{Tr} (\rho P_{\bm{\alpha}}) \ \    \mbox{if}\  \bm{\alpha}\in G, \\
0 \ \ \ \ \ \ \ \ \ \ \ \ \mbox{otherwise},
\end{array}
\right.
\end{eqnarray}
where the first and second lines are due to the linearity and cyclicity of the trace, respectively, the third line is due to the fact that $J_G=J_G^+(\bm{\alpha}) \cup J_G^-(\bm{\alpha})$, with $J_G^+(\bm{\alpha}) := \{\bm{\beta}\in J_G : [P_{\bm{\beta}}',P_{\bm{\alpha}}]=0\}$ and $J_G^-(\bm{\alpha}) := \{\bm{\beta}\in J_G : \{P_{\bm{\beta}}',P_{\bm{\alpha}}\}=0\}$, the fourth line is due to $P_{\bm{\beta}}'^{\dagger}P_{\bm{\beta}}' = P_{\bm{\beta}}^2 = \mathbb{I}$, the fifth line is due to $|J_G|=|J_G^+(\bm{\alpha})|+ |J_G^-(\bm{\alpha})|$, and finally the sixth line is due to the fact that $|J_G^+(\bm{\alpha})|=|J_G|$ when $\bm{\alpha}\in G$ while $|J_G^+(\bm{\alpha})|=|J_G|/2$ otherwise, which in turn is due to Eqs.~(\ref{eq:conditionsonUG}) and the assumption that the matrices $P_{\bm{\beta}}'$ for which $\bm{\beta}\in J_G$ form a set of the form given in Eq.~(\ref{eq:groupuptoaphase}) for some family of matrices $\{P_{\bm{\alpha_i}}\}_{i=1}^{n}$.

An alternative implementation of the above described ENIP can be realised by resorting to the following sequential approach. We begin by considering the $n$ single-qubit local unitaries $\{P_{\bm{\alpha}_i}\}_{i=1}^n$. Then, we fix a sequence of states $\{\rho_{i}\}_{i=0}^n$ defined recursively by
\begin{equation}
\rho_{i} := \frac{1}{2}\left( \rho_{i-1}+P_{\bm{\alpha}_{i-1}}\rho_{i-1}P_{\bm{\alpha}_{i-1}} \right),
\end{equation}
for $i \in \{1,2, \ldots n\}$. This can be achieved physically in each step by having one of the qubit parties flip a coin and classically communicate the result to all the other qubits, with the result of the flip deciding whether the single-qubit local unitary $P_{\bm{\alpha}_{i-1}}$ is applied or not. Then, by setting $\rho_{0} = \rho$ we can easily see that
\begin{equation}\label{Eq:ConvexIteration}
\Pi_G(\rho) = \rho_{n}= \frac{1}{2^n}\sum_{i=1}^{2^{n}} P_{\bm{\beta}_i}' \rho P_{\bm{\beta}_i}'^{\dagger},
\end{equation}
where the matrices $P_{\bm{\beta}_i}'$ are defined in Eq.~(\ref{eq:groupuptoaphase}).

We show in the following a particular realisation of this method for constructing an ENIP for any number of qubits $N$, and hence see how it can be used as a tool within our framework. The identification of alternative ENIPs may proceed by first fixing $N$ and choosing a $G \subset I_{N}$, perhaps based on experimental or physical considerations. One then searches for a family of matrices $\{P_{\bm{\alpha_i}}\}_{i=1}^{n}$ so that Eq.~(\ref{eq:conditionsonUG}) and Eq.~(\ref{eq:groupuptoaphase}) hold. If such a family can be found, then the resultant matrices $\{P_{\bm{\beta}_i}'\}_{i=1}^{2^{n}}$ in Eq.~(\ref{eq:groupuptoaphase}) define a $J_{G}$ that can be used to construct the ENIP according to Eq.~(\ref{eq:groupavarageprojection}). Generally, identifying a valid $G$ and $J_{G}$ can be a difficult task. Nevertheless, it is a process that can be easily automated for small numbers of qubits, where the quantification of multiqubit entanglement still remains an open and relevant problem.

\subsection{Applying the framework}\label{sec:applying_entanglement}

We are now ready to apply the four-step framework introduced in Section \ref{Section:Framework} to the resource theory of multiqubit entanglement. One realisation of this framework has already been achieved in \cite{cianciaruso2016accessible} by considering a fixed ENIP with corresponding EGSs given by the so-called \M3N states $\omega=(\mathbb{I}^{\otimes N}+\sum_{i=1}^3 c_i \sigma_i^{\otimes N})/2^N$ ($N$-qubit mixed states with all maximally mixed marginals), with the $c_{i}\in \mathbb{R}$ constrained so that $\omega$ is positive semidefinite. In the following we will introduce another realisation of our framework based on the so-called \EGN states. As we shall see, the \EGN states allow us to derive lower bounds on multiqubit entanglement that are complementary to those provided by the \M3N states. The steps of our framework for \EGN states are now explained.
%, in the sense that the \EGN states fare particularly well in providing lower bounds in the odd $N$ case, while the \M3N states do so in the even $N$ case.

\begin{description}
\item[Step One: ]{Identify an ENIP and characterise the corresponding set of EGSs.}
\end{description}
In this step we can use the previously discussed method to construct ENIPs and find the corresponding EGSs. By resorting to the following $2(N-1)$ local unitaries
\begin{eqnarray}\label{eq:localunitariesgforNgeq3}
\{P_{\bm{\alpha}_i}\}_{i=1}^{2(N-1)}=\{(\sigma_{3} \otimes \sigma_{3} \otimes \mathbb{I}^{\otimes N-2}),
(\mathbb{I} \otimes \sigma_{3} \otimes \sigma_{3} \otimes \mathbb{I}^{\otimes N-3}),\nonumber \\
\ldots
(\mathbb{I}^{\otimes N-3} \otimes \sigma_{3} \otimes \sigma_{3} \otimes \mathbb{I}),
(\mathbb{I}^{\otimes N-2} \otimes \sigma_{3} \otimes \sigma_{3}) \nonumber \\,
(\sigma_{2} \otimes \sigma_{2} \otimes \mathbb{I}^{\otimes N-2}),
(\mathbb{I} \otimes \sigma_{2} \otimes \sigma_{2} \otimes \mathbb{I}^{\otimes N-3}),\nonumber \\
\ldots
(\mathbb{I}^{\otimes N-3} \otimes \sigma_{2} \otimes \sigma_{2} \otimes \mathbb{I}),
(\mathbb{I}^{\otimes N-2} \otimes \mathbb{I} \otimes \sigma_{2})
\},
\end{eqnarray}
when $N \geq 3$ and to
\begin{equation}\label{eq:localunitariesgforNequal2}
\{P_{\bm{\alpha}_i}\}_{i=1}^{2}= \{(\sigma_{3} \otimes \sigma_{3}), (\mathbb{I} \otimes \sigma_{2})\}
\end{equation}
when $N=2$, as well as to the recursive procedure introduced in Eq.~(\ref{Eq:ConvexIteration}), we identify an ENIP and obtain the family of $N$-qubit EGSs whose matrix representation in the computational basis is given by:
\begin{equation}\label{eq:NqubitEGState}
\varpi = \frac{1}{2^{N}} \left( \mathbb{I}^{\otimes N} + d_{1} \sigma_{1}^{\otimes N-1}\otimes\sigma_2 + d_2 \sigma_{2}^{\otimes N}+ d_3 \sigma_{3}^{\otimes N-1}\otimes\mathbb{I}\right),
\end{equation}
where $d_{1} = {\rm Tr}\left[\varpi (\sigma_{1}^{\otimes N-1}\otimes \sigma_2)\right]$, $d_{2} = {\rm Tr}\left[\varpi \sigma_{2}^{\otimes N}\right]$, and $d_{3} = {\rm Tr}\left[\varpi (\sigma_{3}^{\otimes N-1}\otimes \mathbb{I})\right]$. These states will be referred to as \EGN states and will be denoted in the following also by the triple $\{d_{1},d_{2},d_{3}\}$.

\begin{figure}[t]
    \centering
    \includegraphics[width=0.6\textwidth]{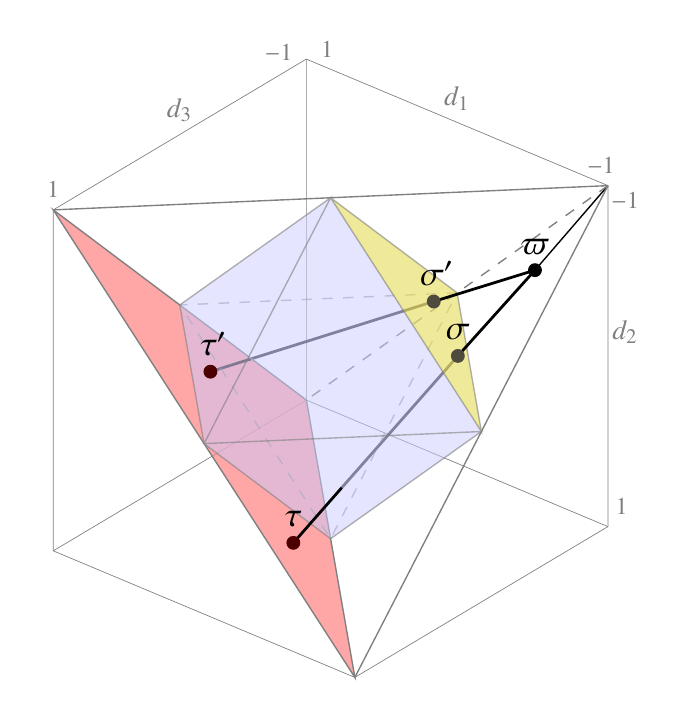}
    \caption{The geometry of \EGN states for odd $N$ and $M > \left \lfloor{N/2}\right \rfloor +1$, together with example choices of optimal states $\sigma$, $\tau$ for the generalised robustness $E_{M}^{R}(\varpi)$. The tetrahedron ${\cal T}_{(-1)^{(N-1)/2}}$ can be seen to contain the $M$-separable octahedron $\mathcal{O}_{1}$ (shaded blue). For any choice of $\varpi$, the closest $M$-separable states $\sigma$ lie in the face of the octahedron closest to $\varpi$ (shaded yellow), and the optimal $\tau$ lie in the base of the tetrahedron (shaded red). The particular choice of the optimal state $\sigma$ is also the closest $M$-separable state with respect to any bona fide distance measure of entanglement. Instead, another optimal state $\sigma'$ can be seen to constitute a valid optimal choice for the standard robustness of $M$-inseparable entanglement since $\tau'$ is also $M$-separable.}
    \label{Fig:Robustness}
\end{figure}

The characterisation of the \EGN states is manifestly different between the odd and even $N$ case.
%When $N=1$, they simply become the states $\varpi=(\mathbb{I}+d_2 \sigma_2)/2$ diagonal in the basis $|\pm_y\rangle = (|0\rangle \pm i |1\rangle)/\sqrt{2}$, which are represented in the $\{d_{1},d_{2},d_{3}\}$-space by the line segment $\mathcal{L}_y$ with end points $\{0,-1,0\}$ and $\{0,1,0\}$.
In the $\{d_{1},d_{2},d_{3}\}$-space, the set of \EGN states with odd $N>1$ is represented by the tetrahedron ${\cal T}_{(-1)^{(N-1)/2}}$ with vertices $\{1,(-1)^{(N-1)/2},1\}$, $\{-1,-(-1)^{(N-1)/2},1\}$, $\{1,-(-1)^{(N-1)/2},-1\}$ and $\{-1,(-1)^{(N-1)/2},-1\}$, see Fig.~\ref{Fig:Robustness}. This tetrahedron is constructed simply by imposing the non-negativity of the four unique eigenvalues of $\varpi$, see \ref{Appendix:CharacterisingVarpi}. Similarly, for even $N$ the set of \EGN states is given in the $\{d_{1},d_{2},d_{3}\}$-space by the unit ball ${\cal B}_{1}$ centred into the origin.

%Consequently, thanks again to the semi-positivity constraint that $\varpi$ has to obey, the set of \EGN states with even $N$ is represented in the $\{d_{1},d_{2},d_{3}\}$-space by the unit ball ${\cal B}_{1}$ centred into the origin.

\begin{description}
\item[Step Two: ]{Characterise the set of free EGSs.}
\end{description}

We now discuss the set $\mathcal{S}_M^{EG_N}$ of $M$-separable \EGN states for any $2\leq M \leq N$. This set can be characterised as the result of applying the fixed ENIP specified in
Eqs.~(\ref{Eq:ConvexIteration}),  (\ref{eq:localunitariesgforNgeq3}) and (\ref{eq:localunitariesgforNequal2}) onto the general set of $M$-separable states given in Eq.~(\ref{Eq:MSep}), see \ref{Appendix:SepEGN} for further details. We see that when $M > \left \lfloor{N/2}\right \rfloor +1$ the $M$-separable \EGN states are such that $|d_1|+|d_2|+|d_3|\leq 1$ and thus fill the set represented in the $\{d_{1},d_{2},d_{3}\}$-space by the unit octahedron ${\cal O}_{1}$ with vertices $\{\pm 1, 0, 0\}$, $\{0, \pm 1, 0\}$ and $\{0, 0, \pm 1\}$, as illustrated in Fig~\ref{Fig:Robustness}. On the other hand, when $M \leq \left \lfloor{N/2}\right \rfloor +1$ all the \EGN states are $M$-separable.

\begin{description}
\item[Step Three: ]{Evaluate $E_{M}(\varpi)$ for all $\varpi \in \mbox{\EGN}$.}
\end{description}

It is immediate to see that $E_{M}(\varpi) =0$ for all $\varpi \in \mbox{\EGN}$ whenever $M \leq \left \lfloor{N/2}\right \rfloor +1$. Therefore, one cannot use the \EGN states to provide non-trivial lower bounds on multiqubit entanglement for such $M$. We instead focus on the cases $M > \left \lfloor{N/2}\right \rfloor +1$, where the $M$-separable states always form the unit octahedron as a strict subset of all \EGN states.

Let us first consider the distance-based measures of $M$-inseparable multiqubit entanglement, where Eq.~(\ref{Eq:DistanceBasedRGS}) shows that we simply need to find the minimal distance from $\varpi$ to the set of \EGN states inside the unit octahedron $\mathcal{O}_{1}$. In the odd $N>1$ case, since all the \EGN states are diagonal in the same basis (\ref{Appendix:CharacterisingVarpi}), we have that distances between them reduce to the corresponding classical distance between the probability distributions formed by their eigenvalues. What is more, since the eigenvalues of the \EGN states with odd $N>1$ are equivalent to the eigenvalues of the \M3N states with even $N$, it happens that the ensuing optimisation problem of classical information geometry has already been solved in \cite{cianciaruso2016accessible}. The result is that, for any choice of distance, the closest $M$-separable \EGN state to $\varpi$ is on the nearest surface of the unit octahedron, with the location specified by the intersection with the extended line connecting $\varpi$ to its corresponding nearest vertex, see Fig.~\ref{Fig:Robustness}. Any bona fide distance-based measure can then be calculated as a monotonically increasing function of the height $h_{\varpi} = \frac{1}{2}\left(\sum_{j=1}^{3} |d_{j}|-1\right) \in [-\frac{1}{2},1]$ above the separable plane (with a negative value indicating that $\varpi$ is in the unit octahedron and hence $M$-separable).

On the other hand, for even $N$, the closest $M$-separable state to $\varpi$ depends on the choice of distance. Nevertheless, since $\sigma_{1}^{\otimes N-1}\otimes\sigma_2$,  $\sigma_{2}^{\otimes N}$ and $\sigma_{3}^{\otimes N-1}\otimes\mathbb{I}$ form a triple of anticommuting matrices, one can easily see that the trace distance between any two \EGN states with even $N$ reduces to (half) the Euclidean distance between their corresponding triples, as is also the case for the \M3N states with odd $N$~\cite{cianciaruso2016accessible}. This means that the trace distance-based measure of $M$-inseparable multiqubit entanglement for $\varpi$ is simply the Euclidean distance from its triple to the unit octahedron.

We now prove that the analytical expression of the robustness of $M$-inseparable multiqubit entanglement $E_{M}^{R}(\varpi)$ of an arbitrary \EGN state $\varpi$ for any odd $N>1$ is the following:
\begin{equation}\label{oddNrobustnessEGN}
E^R_{M}(\varpi) = \left\{
                           \begin{array}{ll}
                             0 \,, & \hbox{$h_\varpi \leq 0$ or $M \leq \left \lfloor{N/2}\right \rfloor + 1$;} \\
                             h_\varpi\,, & \hbox{otherwise.}
                           \end{array}
                         \right.
\end{equation}
To do this, given an \EGN state $\varpi$ and taking into account Eq.~(\ref{Eq:RobustnessRGS}), we need to solve the following simplified optimisation (which can easily be posed as a semidefinite program~\cite{napoli2016robustness})
\begin{equation}\label{Eq:RobustnessEGN}
E_{M}^{R}(\varpi)= \inf_{\tau \in \mbox{\EGN}} \left\lbrace s \geq 0 \left| \frac{\varpi + s \tau}{1+s}=: \sigma \in \mathcal{S}_{M}^{EG_N} \right\rbrace \right. ,
\end{equation}
i.e. we need to find the smallest positive $s$ such that $\sigma=\frac{\varpi + s \tau}{1+s}$ is an $M$-separable \EGN state and $\tau$ is any \EGN state. In other words, we need to prove that $s=h_\varpi$ is the smallest positive $s$ for which $\sigma=\frac{\varpi + s \tau}{1+s}$ is represented in the $\{d_1,d_2,d_3\}$-space by a point belonging to the unit octahedron $\mathcal{O}_1$, and $\tau$ by any point in the tetrahedron ${\cal T}_{(-1)^{(N-1)/2}}$, provided that $h_\varpi \geq 0$ and $M \leq \left \lfloor{N/2}\right \rfloor + 1$, which are the only nontrivial cases where $\varpi$ is not $M$-separable.

In the following we will assume without loss of generality that $\varpi$ belongs to the corner containing the vertex $\{(-1)^{(N-1)/2},(-1)^{(N-1)/2},(-1)^{(N-1)/2}\}$, since all the \EGN states belonging to the other three corners can be obtained from this by simply applying a single-qubit local unitary $\sigma_{i}~\otimes~\mathbb{I}^{\otimes N-1}$, $i \in \{1,2,3\}$, under which any sort of multiqubit entanglement is invariant.

The optimisation in Eq.~(\ref{Eq:RobustnessEGN}) can be solved simply by using the fact that the optimal $\tau$ must be as far from $\varpi$ as possible and is hence represented by a point on the base of the tetrahedron ${\cal T}_{(-1)^{(N-1)/2}}$ with respect to $\varpi$, i.e. given by a triple $\{e_1,e_2,e_3\}$ satisfying
\begin{equation}\label{eq:optimaltaucontraint}
e_1+e_2+e_3=- (-1)^{(N-1)/2},
\end{equation}
shown as the shaded red region in Fig.~\ref{Fig:Robustness}. For a given $\tau$ satisfying this condition, one can then easily see that the optimal $\sigma$ lies on the intersection of the line connecting $\tau$ and $\varpi$ (given by the convex combination $\sigma = \frac{\varpi + s \tau}{1+s}$ for $s \geq 0$) with the face of the unit octahedron $\mathcal{O}_{1}$ closest to $\varpi$, given by any triple $\{s_1,s_2,s_3\}$ satisfying
\begin{equation}\label{eq:optimalsigmaconstraint}
s_1+s_2+s_3=(-1)^{(N-1)/2},
\end{equation}
see again Fig.~\ref{Fig:Robustness} for an illustration. One then finds that $s = h_{\varpi}$, which holds for \emph{any} choice of $\tau$ on the base of the tetrahedron.

It is hence clear that there is not a unique pair of $\tau \in \mbox{\EGN}$ and $\sigma \in \mathcal{S}_{M}^{EG_N}$ satisfying the infimum in Eq.~(\ref{Eq:RobustnessEGN}). We have shown that one can in fact satisfy the infimum with any $\tau$ on the base of the tetrahedron furthest from $\varpi$ and any $\sigma$ on the face of the octahedron closest to $\varpi$, provided that they are colinear with $\varpi$ itself. The optimal $s$ is then given by the plane height $h_{\varpi}$, as shown in Eq.~(\ref{oddNrobustnessEGN}). The non-uniqueness in the optimisation means that the infimum can even be satisfied by an $M$-separable $\tau$ sitting on the face of the octahedron $\mathcal{O}_{1}$ furthest from $\varpi$. A consequence of this is that the robustness of $M$-inseparable multiqubit entanglement $E_{M}^{R}(\varpi)$ of any \EGN state $\varpi$ coincides with the \textit{standard} robustness, where the optimisation over $\tau$ in Eq. \eqref{Eq:RobustnessEGN} is additionally restricted to $M$-separable states~\cite{vidal1999robustness}. The standard robustness was previously calculated for two-qubit Bell-diagonal states in~\cite{akhtarshenas2003robustness}, which have an identical geometry to the odd $N$ \EGN states.

It is also relevant to note that the robustness of $M$-inseparable multiqubit entanglement coincides with (twice) the trace distance-based measure $E^{D_{\text{Tr}}}_{M}(\varpi) = h_{\varpi}/2$ for \EGN states $\varpi$ with odd $N>1$~\cite{cianciaruso2016accessible}. Intriguingly, the closest $M$-separable state to $\varpi$ according to any contractive distance (such as the trace distance) is also a valid $M$-separable \EGN state satisfying the optimisation for the robustness, see Fig.~\ref{Fig:Robustness}. Additionally, we note that the standard and generalised robustness provide, respectively, upper and lower bounds for a family of norms introduced in \cite{regula2017convex} which constitute measures of $M$-inseparable multiqubit entanglement generalising the greatest cross norm \cite{rudolph2001new}. The fact that the two robustness quantifiers coincide in this case then implies that the multiqubit norms are also equal to them for all \EGN states. These simplifications for \EGN states highlight the wide scope of the applicability of our framework to different resource measures when one chooses a suitably simple class of RGSs.

\begin{description}
\item[Step Four: ]{Optimise the lower bound $E_{M}(\Pi(\mathcal{U}_{\otimes}\rho \mathcal{U}_{\otimes}^{\dagger})) \leq E_{M}(\rho)$ over single-qubit local unitaries $\mathcal{U}_{\otimes}$.}
\end{description}

We now consider optimising the lower bound provided through our framework to $E_{M}(\rho)$ for any state $\rho$  by varying over single-qubit unitaries $U_\otimes = \bigotimes_{\alpha=1}^{N} U^{(\alpha)}$ and considering the corresponding \EGN state $\Pi(U_\otimes \rho U_\otimes^\dagger)$, resulting in the maximised lower bound
%We now consider the optimisation of the quantitative lower bound provided by the \EGN state $\Pi(\rho)$ to an arbitrary $N$-qubit state $\rho$ over single-qubit local unitaries applied to the state $\rho$ before the corresponding ENIP $\Pi$, i.e. we consider
\begin{equation}\label{lowopt}
\sup_{U_\otimes} E_{M}(\Pi(U_\otimes \rho U_\otimes^\dagger)) \leq   E_{M}(\rho)\,.
\end{equation}
%where $U_\otimes = \bigotimes_{\alpha=1}^{N} U^{(\alpha)}$ denotes any single-qubit local unitary operation.

Experimentally, the optimised bound can be accessed by measuring a triple of correlations functions $\{\widetilde{d}_j\}$ given by the expectation values of correspondingly rotated Pauli operators on each qubit, $\widetilde{d}_1 = \langle U_\otimes^\dagger (\sigma_1^{\otimes N-1}\otimes \sigma_2) U_\otimes \rangle$, $\widetilde{d}_2 = \langle U_\otimes^\dagger \sigma_2^{\otimes N} U_\otimes \rangle$, $\widetilde{d}_3 = \langle U_\otimes^\dagger (\sigma_3^{\otimes N-1} \otimes \mathbb{I}) U_\otimes \rangle$, and is non-zero whenever $M > \left \lfloor{N/2}\right \rfloor +1$ and $\sum_{j=1}^{3} |\widetilde{d}_j| > 1$. For odd $N>1$, optimality in Eq.~(\ref{lowopt}) for both the family of distance-based measures and the robustness can always be achieved by the choice of $U_\otimes$ such that the quantity $\widetilde{h}_\varpi = \frac12\left( \sum_{j=1}^3 |\widetilde{d}_j|-1\right)$ is maximum. For even $N$, while measures are not generally monotonic functions of $\widetilde{h}_\varpi$, one can take as an ansatz that optimising $\widetilde{h}_\varpi$ provides an improved lower bound.

\begin{table*}[!t]

\centering

\begin{tabular}{ccccc}
 \hline \hline
$N$ & \multicolumn{1}{c}{State} & \multicolumn{1}{c}{$\{\widetilde d_1,\widetilde d_2,\widetilde d_3\}$} & \multicolumn{1}{c}{$\sum_{j=1}^{3} |\widetilde{d}_j|$} & \multicolumn{1}{c}{$\{\theta,\psi,\phi\}$} \\
 \hline
 $3$ &
$|\text{GHZ}^{(3)}\rangle$ & $\left\lbrace 1, -1, 1 \right\rbrace$  &$3$                                       & $\left\lbrace 0,\frac{\pi}{12}, \frac{\pi}{12} \right\rbrace$   \\
%&
%$|\text{W}^{(3)}\rangle $       & $\left\lbrace\frac{1}{\sqrt{3}},-\frac{1}{\sqrt{3}} ,\frac{1}{\sqrt{3}} \right\rbrace$                                                               & $\sqrt{3} $                                                       & $\left\lbrace \cos^{-1}(\frac{1}{\sqrt{3}}),0 ,\frac{\pi}{4} \right\rbrace$ \\

 \hline
  $4$ &
$|\text{GHZ}^{(4)}\rangle$ & $\left\lbrace \frac{1}{\sqrt{2}}, \frac{1}{\sqrt{2}},  0\right\rbrace$ & $\sqrt{2}$                                                                     & $\left\lbrace 0, \frac{\pi}{32}, \frac{\pi}{32} \right\rbrace$   \\
%&
%$|\text{W}^{(4)}\rangle $       & $\left\lbrace\frac{5}{9},\frac{5}{9} ,\frac{5}{9} \right\rbrace$                                                                                           & $\frac{5}{3}$                                                 & $\left\lbrace \cos^{-1}(\frac{1}{\sqrt{3}}),0 ,\frac{\pi}{4} \right\rbrace$ \\

\hline
 $5$ &
$|\text{GHZ}^{(5)}\rangle$ & $\left\lbrace 1, 1,  1\right\rbrace$                                                                                       & $3$                                                         & $\left\lbrace 0, \frac{\pi}{20}, \frac{\pi}{20} \right\rbrace$   \\
%&
%$|\text{W}^{(5)}\rangle $       & $\left\lbrace\frac{7}{9\sqrt{3}},-\frac{7}{9\sqrt{3}} ,\frac{7}{9\sqrt{3}} \right\rbrace$                                                        & $\frac{7}{3\sqrt{3}}$                                         & $\left\lbrace \cos^{-1}(\frac{1}{\sqrt{3}}),0 ,\frac{\pi}{4} \right\rbrace$ \\

 \hline
 $6$ &
$|\text{GHZ}^{(6)}\rangle$ & $\left\lbrace \frac{1}{\sqrt{2}}, - \frac{1}{\sqrt{2}},  0\right\rbrace$ & $\sqrt{2}$                                                         &           $\left\lbrace 0, \frac{\pi}{48}, \frac{\pi}{48} \right\rbrace$   \\
%&
%$\varrho^{(6)}_\text{W}(q) $       & $\left\lbrace0,0 ,-q \right\rbrace$                                                                                                                                                   & $q$                                                                    & $\left\lbrace 0,0, 0 \right\rbrace$ \\

\hline
 $7$ &
$|\text{GHZ}^{(7)}\rangle$ & $\left\lbrace 1, -1,  1\right\rbrace$                                                                                                                                                 & $3$                                                         &           $\left\lbrace 0, \frac{\pi}{28}, \frac{\pi}{28} \right\rbrace$   \\
%&
%$\varrho^{(6)}_\text{W}(q) $       & $\left\lbrace0,0 ,-q \right\rbrace$                                                                                                                                                   & $q$                                                                    & $\left\lbrace 0,0, 0 \right\rbrace$ \\

 \hline
 \hline
 \end{tabular}

     \caption{\label{TableExamples}
Lower bounds to the $M$-inseparable multiqubit entanglement of $|\text{GHZ}^{(N)}\rangle$ when $M > \left \lfloor{N/2}\right \rfloor +1$ can be improved by maximising $\sum_{j=1}^{3} |\widetilde{d}_j| = 2 \widetilde{h}_{\varpi} + 1$, with identical single-qubit unitaries described by the angles $\{\theta,\psi,\phi\}$.}

\end{table*}

\begin{comment}%For me this part is just lengthy and doesn't add much
By using the well known correspondence between the special unitary group ${\sf SU}(2)$ and special orthogonal group ${\sf SO}(3)$, we have that to any one-qubit unitary $U^{(\alpha)}$ corresponds the orthogonal $3\times 3$ matrix $O^{(\alpha)}$ such that
\begin{equation}
U^{(\alpha)} \vec{n}\cdot\vec{\sigma}U^{(\alpha)\dagger} = (O^{(\alpha)} \vec{n})\cdot\vec{\sigma},
\end{equation}
where $\vec{n}=\{n_1,n_2,n_3\}\in\mathbb{R}^3$ and $\vec{\sigma} = \{\sigma_1,\sigma_2,\sigma_3\}$ is the vector of Pauli matrices. We have then that
\begin{equation}
\sup_{\{U^{(\alpha)}\}} (|\widetilde{d}_1|+|\widetilde{d}_2|+|\widetilde{d}_3|)=\sup_{\{O^{(\alpha)}\}} (|\widetilde{T}_{11\cdots 2}|+|\widetilde{T}_{22\cdots 2}|+|\widetilde{T}_{33\cdots 0}|),
\end{equation}
where
\begin{eqnarray}
\widetilde{T}_{i_1i_2\cdots i_N} &=&  \sum_{j_1j_2\cdots j_N} T_{j_1j_2\cdots j_N}O^{(1)}_{i_1j_1}O^{(2)}_{i_2j_2}\cdots O^{(N)}_{i_Nj_N} \\
\widetilde{T}_{i_1i_2\cdots i_{N-1} 0} &=&  \sum_{j_1j_2\cdots j_{N-1}} T_{j_1j_2\cdots j_{N-1} 0} O^{(1)}_{i_1j_1}O^{(2)}_{i_2j_2}\cdots O^{(N-1)}_{i_{N-1}j_{N-1}}
\end{eqnarray}
and
\begin{equation}
{T}_{i_1i_2\cdots i_N}= \text{Tr}\left[\rho \left(\sigma_{i_1}\otimes\sigma_{i_2}\otimes\cdots\otimes\sigma_{i_N}\right) \right].
\end{equation}
\end{comment}

In Table \ref{TableExamples} we can see how useful our results are on the paradigmatic example of the $N$-qubit GHZ state~\cite{greenberger1990bell}
\begin{equation}
|\text{GHZ}^{(N)}\rangle = \frac{1}{\sqrt{2}} \left(|00\cdots 00 \rangle + |11\cdots 11 \rangle\right),
\end{equation}
with $N\geq 3$, which constitutes a primary resource for quantum computation and metrology~\cite{giovannetti2011advances}. See Ref.~\cite{cianciaruso2016accessible} for a comparison to results for \M3N states. Here, due to the qubit permutation invariance of $|\text{GHZ}^{(N)}\rangle$\cite{cianciaruso2016accessible}, optimisation of $\widetilde{h}_{\varpi}$ can be achieved by setting all the single-qubit unitaries to be identical, i.e. $U^{(\alpha)} = U_{2}$ for all $\alpha$. Here, $U_{2}$ can be parameterised by $3$ angles $\{\theta, \psi, \phi\}$ in the following way,
\begin{equation}
U_2 =  \left( \begin{array}{cc}
\cos\frac{\theta}{2} e^{-i\frac{\psi+\phi}{2}} & -i \sin\frac{\theta}{2} e^{-i\frac{\phi-\psi}{2}}  \\
-i \sin\frac{\theta}{2} e^{i\frac{\phi-\psi}{2}} & \cos\frac{\theta}{2} e^{i\frac{\psi+\phi}{2}}   \\
 \end{array} \right).
\end{equation}

%turns out to be achieved when $O^{(1)}=O^{(2)}=\cdots= O^{(N)}$, i.e. by performing the maximisation over just the three angles $\{\theta, \psi, \phi\}$ which determine the orthogonal matrix $O^{(\alpha)}$ corresponding to an arbitrary single-qubit unitary $$U^{(\alpha)} =  \left( \begin{array}{cc}
%\cos\frac{\theta}{2} e^{-i\frac{\psi+\phi}{2}} & -i \sin\frac{\theta}{2} e^{-i\frac{\phi-\psi}{2}}  \\
%-i \sin\frac{\theta}{2} e^{i\frac{\phi-\psi}{2}} & \cos\frac{\theta}{2} e^{i\frac{\psi+\phi}{2}}   \\
% \end{array} \right).$$

\section{Discussion}\label{Section:Discussion}

% For me I always found the following a bit tenuous, as we require evaluation of particular instances (e.g. the relative entropy) to see the tightness, whereas here we focus on the robustness more, for which I could not find a calculation of the global robustness entanglement .......
%??We note that for any odd $N>1$ GHZ state, our optimised quantitative bound specified through Table~\ref{TableExamples} does not only reach the maximum $h_\varpi=1$ but becomes tight as well, thus returning the exact value of their global multiqubit entanglement via Eq.~(\ref{evNpartindep}), despite the fact that such states are not (and are very different from) \EGN states. Interestingly, this result complements what has been found in \cite{cianciaruso2016accessible}, i.e. that the quantitative lower bound provided by the \M3N states is tight for even $N$, while it does not compare as nicely for any odd $N$.??

Our general framework provides a clearcut approach to finding lower bounds to resource measures evaluated on arbitrary states. This framework is founded upon the hereby introduced concepts of resource non-increasing projections and the corresponding resource guarantor states. Each step in the framework is feasible to carry out. The first step can be performed by systematically identifying an RNIP, as we have shown in Sec.~\ref{Sec:Systematic}, or by using intuition about the resource under consideration, as may be done for coherence. The second step can be realised by characterising the intersection between free states and RGSs, as we have shown in \ref{Appendix:SepEGN} for multiqubit entanglement. The resultant optimisation in step three is necessarily simpler than for the corresponding arbitrary state due to the simplified structure of the RGS. We have furthermore shown that the optimisation can be expressed as an SDP for the resource robustness and can hence be evaluated computationally with little overhead. Finally, varying over local unitaries in the fourth step can be a restricted optimisation over a constrained and/or discrete set of candidates. Moreover, our framework is more experimentally friendly in the sense that it necessarily requires fewer measurements than a full state tomography.

We illustrated the relevance of this framework for multiqubit entanglement by constructing a general accessible formalism to identify entanglement non-increasing projections and their resource guarantor states, giving a particular example of a projection that results in suitably defined \EGN states. We then proceeded to complete the steps of our framework for this example, allowing us to find analytic lower bounds to the multiqubit entanglement of GHZ states that can be measured experimentally using only three local measurement settings.

Our approach can be understood as a particular type of quantitative resource witness~\cite{eisert2007quantitative,guhne2007estimating}, providing an approximation of the resources present in a system based on the results of a limited selection of measurement settings. It will be of further interest to compare the efficiency of lower bounds arising from our framework to those arising from other approaches, as has been done specifically for entanglement in~\cite{cianciaruso2016accessible}. Nevertheless, our framework relies on the universal concept of resource monotonicity, and can hence be applied in principle to a vast range of possible resource measures.

We have focused in this work on the provision of lower bounds to resource measures. These lower bounds are useful for verifying the minimum usefulness of a resource state. In practice, whenever a measure can be linked to the performance of an operational protocol, our lower bounds can be harnessed to guarantee a worst-case performance of using a given resource state. Nevertheless, evaluating \emph{upper} bounds on relevant resource measures is also important, allowing for better comparison between resource states and hence a finer grained identification of states most useful in an operational setting. Our framework is geared towards providing lower bounds by contracting the state space using resource non-increasing projections. It will be of future interest to identify dual frameworks able to identify upper bounds for a given class of resource states.

By applying our framework to \EGN states, we have been able to provide new results for evaluating the robustness of entanglement in both \EGN states with odd $N$ and  \M3N states with even $N$, complementing previous evaluations of the robustness of entanglement for two-qubit Bell diagonal states~\cite{akhtarshenas2003robustness}. Our results show that the robustness coincides, for Bell diagonal states, with the plane height $h_{\varpi}$, which equates the two-qubit concurrence and half the trace distance-based measure of entanglement. Our approach therefore allowed the evaluation of the robustness of entanglement, which is an NP-hard problem~\cite{brandao2005quantifying}, to be simplified to an intuitive geometric optimisation for relevant classes of states. It is hoped that our framework will provide further simplifications when using alternative resource non-increasing projections.

Quantum resources embody the power behind presently developing quantum technologies. These technologies will require rigorous verification, through benchmarking, of the resources present in the employed devices. Our framework allows for a quantitative benchmark with a low overhead. Some of the next steps of our work could be to provide a variety of new lower bounds to resource measures stemming from different choices of projections. From an analytical perspective, it will be of interest to formalise whether a link exists between the strength of the projection (i.e.~the amount of resource lost) and the simplicity of the corresponding family of resource guarantor states. Here it is expected that one necessarily loses a lot of resource by projecting onto a simple family. On the other hand, experimentally it will be interesting to harness our established lower bounds for concrete applications and proof-of-concept experiments to verify and quantify the resources present in complex quantum systems.% We hope to begin future collaboration with experimentalists to emphasise the usefulness of our framework.

\ack{
\vspace*{-.3cm}
We thank Otfried G\"uhne and Nicolangelo Iannella for informative discussions, as well as anonymous referees for crucial feedback and comments. We acknowledge funding from the European Research Council (ERC) under the Starting Grant GQCOP (Grant No.~637352) and the Foundational Questions Institute (fqxi.org) under the Physics of the Observer Programme (Grant No.~FQXi-RFP-1601).

\appendix

\section{Properties of tensor products of Pauli matrices}\label{Appendix:Palpha}

For any $\bm{\alpha},\bm{\beta}\in I_N$, it holds that $P_{\bm{\alpha}}^2 = \mathbb{I}$ and that $P_{\bm{\alpha}}$ and $P_{\bm{\beta}}$ can only either commute (i.e., $[P_{\bm{\alpha}},P_{\bm{\beta}}]=0$) or anti-commute (i.e., $\{P_{\bm{\alpha}},P_{\bm{\beta}}\}=0$). More specifically, $[P_{\bm{\alpha}},P_{\bm{\beta}}]=0$ if there is an even number (including zero) of indices $i\in\{1,\cdots,N\}$ such that $\{\sigma_{\alpha_i},\sigma_{\beta_i}\}=0$, whereas $\{P_{\bm{\alpha}},P_{\bm{\beta}}\}=0$ otherwise. Moreover, given any set of matrices $\{P_{\bm{\alpha_i}}\}_{i=1}^{n}$, we have that $P_{\bm{\alpha_1}}P_{\bm{\alpha_2}}\cdots P_{\bm{\alpha_n}}$ is equal to either $\pm P_{\bm{\beta}}$ or $\pm i P_{\bm{\beta}}$ for some $\bm{\beta}\in I_N$. In the following, with abuse of notation, we will denote any of the matrices $\pm P_{\bm{\beta}}$ or $\pm i P_{\bm{\beta}}$ simply by $P_{\bm{\beta}}'$ as their unitary transformation on any state $\rho$, i.e. $P_{\bm{\beta}}' \rho P_{\bm{\beta}}'^{\dagger}$, provides exactly the same output. Finally, given an arbitrary matrix $P_{\bm{\alpha}}$, we get that $[P_{\bm{\alpha}},P_{\bm{\alpha_1}}P_{\bm{\alpha_2}}\cdots P_{\bm{\alpha_n}}]=0$ if $P_{\bm{\alpha}}$  anti-commutes with an even number (including zero) of matrices $\{P_{\bm{\alpha_i}}\}_{i=1}^{n}$, while $\{P_{\bm{\alpha}},P_{\bm{\alpha_1}}P_{\bm{\alpha_2}}\cdots P_{\bm{\alpha_n}}\}=0$ otherwise. As a consequence, if one considers the following set composed of $2^n$ matrices:
\begin{equation}
\{P_{\bm{\beta}_i}'\}_{i=1}^{2^{n}}=
\left \{
  \begin{tabular}{c}
  $\mathbb{I}^{\otimes N}$ \\
  $\{P_{\bm{\alpha}_{i_1}}\}_{i_{1}=1}^{n}$ \\
  $\{P_{\bm{\alpha}_{i_2}}P_{\bm{\alpha}_{i_1}}\}_{i_{2}>i_{1}=1}^{n}$ \\
  $\cdots$ \\
  $\{P_{\bm{\alpha}_{i_n}} \ldots P_{\bm{\alpha}_{i_2}}P_{\bm{\alpha}_{i_1}}\}_{i_{n}>\ldots>i_{2}>i_{1}=1}^{n}$ \\
  \end{tabular}
\right \}, \nonumber
\end{equation}
it happens that an arbitrary matrix $P_{\bm{\alpha}}$ can only either commute with all the above listed matrices $P_{\bm{\beta}_i}'$ or commute with half of them and anti-commute with the remaining half.

\section{Eigendecomposition of the \texorpdfstring{\EGN}{EGN} states}\label{Appendix:CharacterisingVarpi}

For odd $N>1$, the \EGN states are all diagonal in the following basis:
\begin{equation}\label{eq:EigenvectorsEGNstatesNodd}
|\beta_j^{\pm}\rangle=\frac{1}{\sqrt{2}}\left(\mathbb{I}^{\otimes N} \pm i \sigma_1^{\otimes N} \right) |j\rangle,
\end{equation}
with the corresponding eigenvalues given by:
\begin{equation}\label{eq:EigenvaluesEGNstatesNodd}
\lambda_{p,q}^{\pm}=\frac{1}{2^N}\left[1\pm (-1)^q d_1 \pm (-1)^{(N-1)/2}(-1)^p d_2 +(-1)^{p-q} d_3 \right],
\end{equation}
where $i$ is the imaginary unit, $j\in \{1,\cdots,2^{N-1}\}$, $\{|j\rangle\}_{i=1}^{2^N}$ is the binary ordered $N$-qubit computational basis, while $p$ and $q$ are defined as
\begin{eqnarray}\label{eq:definitionofglobalparity}
\sigma_3^{\otimes N}|\beta_i^{\pm}\rangle &=& (-1)^p |\beta_i^{\pm}\rangle, \\
\mathbb{I}^{\otimes N-1}\otimes\sigma_3|\beta_i^{\pm}\rangle &=& (-1)^q |\beta_i^{\pm}\rangle. \label{eq:definitionofNthqubitparity}
\end{eqnarray}
For even $N$, the eigenvalues of the \EGN states are given by
\begin{equation}\label{eq:EigenvaluesEGNStatesNodd}
\lambda_{\pm}=\frac{1}{2^N}\left(1\pm \sqrt{d_1^2+d_2^2+d_3^2}\right),
\end{equation}
as it can be easily shown by using the fact that in this case $\sigma_{1}^{\otimes N-1}\otimes\sigma_2$,  $\sigma_{2}^{\otimes N}$ and $\sigma_{3}^{\otimes N-1}\otimes\mathbb{I}$ form a triple of anticommuting matrices.

\section{Characterising the $M$-separable \texorpdfstring{\EGN}{EGN} states}\label{Appendix:SepEGN}

Here we characterise the set $\mathcal{S}_M^{EG_N}$ of $M$-separable \EGN states for any $2\leq M \leq N$.  We see that for $M > \left \lfloor{N/2}\right \rfloor +1$, all and only the $M$-separable \EGN states satisfy $|d_1|+|d_2|+|d_3|\leq 1$, and hence fill the set represented in the $\{d_{1},d_{2},d_{3}\}$-space by the unit octahedron ${\cal O}_{1}$ with vertices $\{\pm 1, 0, 0\}$, $\{0, \pm 1, 0\}$ and $\{0, 0, \pm 1\}$. Instead, when $M \leq \left \lfloor{N/2}\right \rfloor +1$, all the \EGN states are $M$-separable.

To prove this, it will be useful to introduce a notion of separability that depends on a particular partition of the composite system under consideration, as opposed to the already introduced notion of $M$-separability, which rather considers indiscriminately all the partitions with a set number $M$ of parties. In order to characterise the possible partitions of an $N$-qubit system, we will employ the following notation \cite{blasone2008hierarchies}:
\begin{itemize}
\item the positive integer $M$, $2\leq M \leq N$, representing the number of subsystems;
\item the set of positive integers $\{K_\alpha\}_{\alpha=1}^M=\{K_1, K_2,\cdots, K_M\}$, where a given $K_\alpha$ represents the number of parties belonging to the $\alpha$-th subsystem;
\item the set of sequences of positive integers $\{Q_\alpha\}_{\alpha=1}^M$, with $Q_\alpha =\left\lbrace  i_1^{(\alpha)},i_2^{(\alpha)},\cdots,i_{K_\alpha}^{(\alpha)}\right\rbrace $, $i_j^{(\alpha)}\in \{1,\cdots,N\}$ and $Q_\alpha\cap Q_{\alpha'}=\varnothing$ for $\alpha\neq \alpha'$, where a given sequence $Q_\alpha$ represents precisely the parties belonging to the $\alpha$-th subsystem.
\end{itemize}
In the following, $\widetilde{Q}_M := \{Q_\alpha\}_{\alpha=1}^M$ will denote a generic $M$-partition of an $N$-qubit system and we will assume without loss of generality that $N\in Q_M$, i.e. the $N$-th qubit is always contained in the $M$-th subsystem. The separable states with respect to the $M$-partition $\widetilde{Q}_M$ are then defined as states $\varsigma$ of the form
\begin{equation}\label{Eq:SeparableStates}
\varsigma = \sum_{i} p_{i}\  \tau_{i}^{(1)} \otimes \tau_{i}^{(2)} \otimes \ldots \otimes \tau_{i}^{(M)},
\end{equation}
where $\{p_{i}\}$ is a probability distribution and $\tau_{i}^{(\alpha)}$ are arbitrary states of the  $\alpha$-th subsystem. In other words, any $\widetilde{Q}_M$-separable state can be written as a convex combination of product states that are all factorised with respect to the same partition $\widetilde{Q}_M$. The set of $\widetilde{Q}_M$-separable states will be denoted as  $\mathcal{S}_{\widetilde{Q}_M}$.

To achieve the characterisation of the $M$-separable \EGN states, we first need to identify the sets $\mathcal{S}_{\widetilde{Q}_M}^{EG_N}$ of $\widetilde{Q}_M$-separable \EGN states obtained by considering all the possible $M$-partitions  $\widetilde{Q}_M$, being the convex hull of their union exactly the set $\mathcal{S}_{M}^{EG_N}$, i.e.
\begin{equation}\label{eq:MsepequalconvexhullunionQalphasep}
\mathcal{S}_{M}^{EG_N} = \conv\left( \bigcup_{\widetilde{Q}_M}\mathcal{S}_{\widetilde{Q}_M}^{EG_N} \right).
\end{equation}

Moreover, we will also need to use the fact that, on one hand, the set of the triples $\{c_{1},c_{2},c_{3}\}$, with $c_{i} = \mbox{{\rm Tr}}(\rho \sigma_{i}^{\otimes N})$, obtained by considering any possible $N$-qubit state $\rho$ is~\cite{cianciaruso2016accessible}
\begin{itemize}
\item{the unit ball ${\cal B}_{1}$, when $N$ is odd;}
\item{the tetrahedron ${\cal T}_{(-1)^{N/2}}$, when $N$ is even.}
\end{itemize}
On the other hand, as we have seen here, the set of the triples $\{d_{1},d_{2},d_{3}\}$, with $d_{1} = {\rm Tr}\left[\rho (\sigma_{1}^{\otimes N-1}\otimes \sigma_2)\right]$, $d_{2} = {\rm Tr}\left[\rho \sigma_{2}^{\otimes N}\right]$, and $d_{3} = {\rm Tr}\left[\rho (\sigma_{3}^{\otimes N-1}\otimes \mathbb{I})\right]$, obtained by considering any possible $N$-qubit state $\rho$ is
\begin{itemize}
\item{the unit ball ${\cal B}_{1}$, when $N$ is even;}
\item{the line segment ${\cal L}_{1}:=\{(t,t,1) \ | -1 \leq t \leq 1\}$, when $N=1$;}
\item{the tetrahedron ${\cal T}_{(-1)^{(N-1)/2}}$, when $N>1$ is odd.}
\end{itemize}
Herein, we shall refer to the triple $\{c_{1},c_{2},c_{3}\}$, with $c_{i} = \mbox{{\rm Tr}}(\rho \sigma_{i}^{\otimes N})$, as the \M3N triple of the state $\rho$. On the other hand, the triple $\{d_{1},d_{2},d_{3}\}$, with $d_{1} = {\rm Tr}\left[\rho (\sigma_{1}^{\otimes N-1}\otimes \sigma_2)\right]$, $d_{2} = {\rm Tr}\left[\rho \sigma_{2}^{\otimes N}\right]$, and $d_{3} = {\rm Tr}\left[\rho (\sigma_{3}^{\otimes N-1}\otimes \mathbb{I})\right]$, will be referred to as \EGN triple of the state $\rho$.

Now each set $\mathcal{S}_{\widetilde{Q}_M}^{EG_N}$ coincides with the set $\Pi_G \left[ \mathcal{S}_{\widetilde{Q}_M} \right]$ obtained by projecting all the $\widetilde{Q}_M$-separable states onto the \EGN states via the ENIP $\Pi_G$ defined in Eqs.~(\ref{Eq:ConvexIteration}),  (\ref{eq:localunitariesgforNgeq3}) and (\ref{eq:localunitariesgforNequal2}). Therefore,  the $\widetilde{Q}_M$-separable \EGN states are represented in the $\{d_{1},d_{2},d_{3}\}$-space by the \EGN triples $\{s_1,s_2,s_3\}$, $s_{1} = {\rm Tr}\left[\varsigma (\sigma_{1}^{\otimes N-1}\otimes \sigma_2)\right]$, $s_{2} = {\rm Tr}\left[\varsigma \sigma_{2}^{\otimes N}\right]$, $s_{3} = {\rm Tr}\left[\varsigma (\sigma_{3}^{\otimes N-1}\otimes \mathbb{I})\right]$, corresponding to all the elements $\varsigma$ of $\mathcal{S}_{\widetilde{Q}_M}$. These are given by
\begin{equation}\label{Eq:SeperableEGNification}
s_j = \sum_{i} p_{i} \prod_{\alpha =1}^{M-1} c_{i,j}^{(\alpha)} d_{i,j}^{(M)},
\end{equation}
where we denote $c_{i,j}^{(\alpha)} = \mbox{{\rm Tr}} \left(\tau_{i}^{(\alpha)} \sigma_{j}^{\otimes K_{\alpha}}\right)$ as the $j$-th component of the \MKA triple $\vec{c}_{i}^{(\alpha)}=\{c_{i,1}^{(\alpha)},c_{i,2}^{(\alpha)},c_{i,3}^{(\alpha)}\}$ corresponding to the arbitrary state $\tau_{i}^{(\alpha)}$ of the $\alpha$-th subsystem, with $\alpha < M$, while $d_{i,j}^{(M)} = \mbox{{\rm Tr}} \left(\tau_{i}^{(\alpha)} \sigma_{j}^{\otimes K_{M}}\right)$ is the $j$-th component of the \EGKM triple $\vec{d}_{i}^{(M)}=\{d_{i,1}^{(M)},d_{i,2}^{(M)},d_{i,3}^{(M)}\}$ corresponding to the arbitrary state $\tau_{i}^{(M)}$ of the $M$-th subsystem (which contains the $N$-th qubit).
Eq.~(\ref{Eq:SeperableEGNification}) can be easily proved by resorting to Eq.~(\ref{Eq:SeparableStates}). For example, when considering the case $j=1$, we get:
\begin{eqnarray}
s_{1} &=& \mbox{{\rm Tr}}\left[ \varsigma (\sigma_{1}^{\otimes N-1} \otimes \sigma_2)\right] \nonumber \\
&=& \mbox{{\rm Tr}} \left[ \left( \sum_{i} p_{i} \tau_{i}^{(1)} \otimes \ldots \otimes \tau_{i}^{(M-1)} \otimes \tau_{i}^{(M)} \right) (\sigma_{1}^{\otimes N-1} \otimes \sigma_2)\right] \nonumber \\
&=&\sum_{i} p_{i} \mbox{{\rm Tr}}\left[ \tau_{i}^{(1)} \sigma_{1}^{\otimes K_{1}} \otimes \ldots \otimes \tau_{i}^{(M-1)} \sigma_{1}^{\otimes K_{M-1}}\otimes \tau_i^{(M)}(\sigma_{1}^{\otimes K_{M}-1}\otimes\sigma_{2}) \right] \nonumber \\
&=&\sum_{i} p_{i} \prod_{\alpha =1}^{M-1} \mbox{{\rm Tr}} \left( \tau_{i}^{(\alpha)} \sigma_{1}^{\otimes K_{\alpha}} \right) \mbox{{\rm Tr}} \left[ \tau_{i}^{(M)} (\sigma_{1}^{\otimes K_{M}-1}\otimes \sigma_2) \right] \nonumber \\
&=& \sum_{i} p_{i} \prod_{\alpha =1}^{M-1} c_{i,1}^{(\alpha)} d_{i,1}^{(M)}.
\end{eqnarray}

Eq.~(\ref{Eq:SeperableEGNification}) can be simplified further by introducing the Hadamard product as the componentwise multiplication of vectors, i.e.~for $\vec{u}=\{u_{1},u_{2},u_{3}\}$ and $\vec{v}=\{v_{1},v_{2},v_{3}\}$ the Hadamard product is $\vec{u} \circ \vec{v}= \{u_{1}v_{1},u_{2}v_{2},u_{3}v_{3}\}$. Using the Hadamard product gives
Eq.~(\ref{Eq:SeperableEGNification}) as
\begin{equation}\label{Eq:SVector}
\vec{s} = \sum_{i} p_{i} \vec{c}_{i}^{(1)}  \circ \ldots \circ \vec{c}_{i}^{(M-1)} \circ \vec{d}_{i}^{(M)},
\end{equation}
i.e., that the \EGN triple of any $\widetilde{Q}_M$-separable state is a convex combination of the Hadamard products between the \MKA triples corresponding to the first $M-1$ subsystem states and the \EGKM triple of the $M$-th subsystem state.

By assuming that the $N$-th qubit belongs to the $M$-th subsystem, we get that $\vec{c}_{i}^{(\alpha)} \in {\cal B}_{1}$ when $\alpha<M$ and $K_{\alpha}$ is odd, $\vec{c}_{i}^{(\alpha)} \in {\cal T}_{(-1)^{K_{\alpha}/2}}$ when $\alpha<M$ and $K_{\alpha}$ is even, $\vec{d}_{i}^{(M)} \in {\cal B}_{1}$ when $K_{M}$ is even, $\vec{d}_{i}^{(M)} \in {\cal L}_{1}$ when $K_{M}=1$, and finally $\vec{d}_{i}^{(M)} \in {\cal T}_{(-1)^{(K_{M}-1)/2}}$ when $K_{M}>1$ is odd. As a consequence, $\mathcal{S}_{\widetilde{Q}_M}^{EG_N}$ is represented by the following set
\begin{equation}\label{Eq:SeparableM3NHadamard}
\mathcal{S}_{\widetilde{Q}_M}^{EG_N} = \conv \left( A^{(1)} \circ \ldots \circ A^{(M-1)} \circ A^{(M)} \right),
\end{equation}
with
\begin{equation}
A^{(\alpha)} =\left\{
  \begin{array}{lr}
    {\cal B}_{1} \,\,\,\,\,\,\,\,\,\,\,\,\,\,\,\,\,\,\,\,\,\,\,\,\,\,\,\,\,\, \mbox{if}\  \alpha<M,\ K_{\alpha} \mbox{ is odd}\\
    {\cal T}_{(-1)^{K_{\alpha}/2}} \,\,\,\,\,\,\,\,\,\,\,\, \mbox{if } \alpha<M,\  K_{\alpha} \mbox{ is even,}\\
    {\cal B}_{1} \,\,\,\,\,\,\,\,\,\,\,\,\,\,\,\,\,\,\,\,\,\,\,\,\,\,\,\,\,\,\mbox{if } \alpha=M,\   K_{M} \mbox{ is even,} \\
    {\cal L}_{1} \,\,\,\,\,\,\,\,\,\,\,\,\,\,\,\,\,\,\,\,\,\,\,\,\,\,\,\,\,\, \mbox{if } \alpha=M,\   K_{M}= 1 \\
    {\cal T}_{(-1)^{(K_{\alpha}-1)/2}} \,\,\,\, \mbox{if } \alpha=M,\  K_{M}>1 \mbox{ is odd,}\\

  \end{array}
\right.
\end{equation}
where we define the Hadamard product between any two sets $A$ and $B$ as
\begin{equation}
A \circ B = \{\vec{a} \circ \vec{b} \, | \, \vec{a}\in A \, , \, \vec{b} \in B\}.
\end{equation}
The commutativity and associativity of the Hadamard product allow us to rearrange the ordering in Eq.~(\ref{Eq:SeparableM3NHadamard}) in the following way
\begin{equation}\label{Eq:SeparableEGNSimpleHadamard}
\mathcal{S}_{\widetilde{Q}_M}^{EG_N} =\left\{
  \begin{array}{lr}
       \conv \left(A^{(1,\cdots,M-1)}  \circ \mathcal{B}_1 \right) \,\,\,\,\,\,\,\,\,\,\,\,\,\,\,\,\,\,\,\,\,\,\,\,\,\,\,\,\,\,\mbox{if } K_{M} \mbox{ is even,} \\
    \conv \left(A^{(1,\cdots,M-1)}  \circ \mathcal{L}_1 \right) \,\,\,\,\,\,\,\,\,\,\,\,\,\,\,\,\,\,\,\,\,\,\,\,\,\,\,\,\,\, \mbox{if } K_{M}= 1 \\
    \conv \left(A^{(1,\cdots,M-1)}  \circ {\cal T}_{(-1)^{(K_{M}-1)/2}} \right) \, \mbox{if } K_{M}>1 \mbox{ is odd,}\\

  \end{array}
\right.
\end{equation}
where
\begin{equation}\label{eq:hadamardproducoffirstMminus1subsystems}
A^{(1,\cdots,M-1)}:= \bigcirc_{\alpha=1}^{M-1} A^{(\alpha)} = \left( \mathop{\mathlarger{\mathlarger{\mathlarger{\bigcirc}}}}_{\mathsmaller{\mu<M: K_{\mu} \even}}{\cal T}_{(-1)^{K_{\mu}/2}}\right) \circ \left( \mathop{\mathlarger{\mathlarger{\mathlarger{\bigcirc}}}}_{\mathsmaller{\nu<M: K_{\nu} \odd}}{\cal B}_{1}\right)
\end{equation}
and $\bigcirc_{\alpha=1}^{n} A^{(\alpha)} := A^{(1)} \circ A^{(2)} \circ \ldots \circ A^{(n)}$.

Now by using Eq.~(\ref{Eq:SeparableEGNSimpleHadamard}) together with the following equations \cite{cianciaruso2016accessible}
\begin{eqnarray}\label{Eq:HadamardProducts}
{\cal T}_{-1} \circ {\cal T}_{-1}  &=& {\cal T}_{1}, \nonumber \\
{\cal T}_{1} \circ {\cal T}_{1}    &=& {\cal T}_{1}, \nonumber \\
{\cal T}_{1} \circ {\cal T}_{-1}   &=& {\cal T}_{-1}, \nonumber \\
{\cal T}_{\pm 1} \circ {\cal B}_{1} &=& {\cal B}_{1},  \\
{\cal T}_{\pm 1} \circ {\cal L}_{1} &=& {\cal T}_{\pm 1}, \nonumber \\
{\cal B}_{ 1} \circ {\cal L}_{1} &=& {\cal B}_{ 1}, \nonumber \\
\conv \left( \bigcirc_{\mathsmaller{i=1}}^{n} {\cal B}_{1} \right) &=& {\cal O}_{1}\ \ \forall n \geq 2. \nonumber
\end{eqnarray}
and the fact that $\conv(A)=A$ if $A$ is convex, we identify the following cases:
\begin{enumerate}
\item{if $K_{\alpha}$ is even for any $\alpha$ then
\begin{eqnarray}
\mathcal{S}_{\widetilde{Q}_M}^{EG_N} &=& \conv \left[ \left( \mathop{\mathlarger{\mathlarger{\mathlarger{\bigcirc}}}}_{\mathsmaller{\mu < M: K_{\mu} \even}}{\cal T}_{(-1)^{K_{\mu}/2}}\right) \circ {\cal B}_{1} \right] \nonumber \\
&=& \conv \left( {\cal T}_{\pm 1} \circ {\cal B}_{1}\right) \nonumber \\
&=& \conv \left( {\cal B}_{1} \right) \nonumber \\
&=& {\cal B}_{1};
\end{eqnarray}\label{Condition:AllEven}
}

\item{if $K_{\alpha}$ is even for all values of $\alpha<M$ and $K_M=1$ then
\begin{eqnarray}
\mathcal{S}_{\widetilde{Q}_M}^{EG_N} &=& \conv \left[\left( \mathop{\mathlarger{\mathlarger{\mathlarger{\bigcirc}}}}_{\mathsmaller{\mu < M: K_{\mu} \even}}{\cal T}_{(-1)^{K_{\mu}/2}} \right) \circ {\cal L}_{1}\right] \nonumber \\
&=& \conv \left( {\cal T}_{(-1)^{(N-1)/2}} \circ {\cal L}_{1}\right) \nonumber \\
&=& {\cal T}_{(-1)^{(N-1)/2}};
\end{eqnarray}
\label{Condition:AllEvenExceptKMequal1}
}

\item{if $K_{\alpha}$ is even for all values of $\alpha<M$ and $K_M>1$ is odd then
\begin{eqnarray}
\mathcal{S}_{\widetilde{Q}_M}^{EG_N} &=& \conv \left[\left( \mathop{\mathlarger{\mathlarger{\mathlarger{\bigcirc}}}}_{\mathsmaller{\mu < M: K_{\mu} \even}}{\cal T}_{(-1)^{K_{\mu}/2}} \right) \circ  {\cal T}_{(-1)^{(K_M-1)/2}}\right] \nonumber \\
&=& \conv \left( {\cal T}_{(-1)^{(N-K_M)/2}} \circ {\cal T}_{(-1)^{(K_M-1)/2}}\right) \nonumber \\
&=& {\cal T}_{(-1)^{(N-1)/2}};
\end{eqnarray}
\label{Condition:AllEvenExceptKMgeq1Odd}
}

\item{if $K_{\alpha}$ is odd for just one value of $\alpha<M$ then
\begin{eqnarray}
\mathcal{S}_{\widetilde{Q}_M}^{EG_N} &=& \conv \left[\left( \mathop{\mathlarger{\mathlarger{\mathlarger{\bigcirc}}}}_{\mathsmaller{\mu < M: K_{\mu} \even}}{\cal T}_{(-1)^{K_{\mu}/2}} \right) \circ {\cal B}_{1} \circ {\cal B}_{1} \right] \nonumber \\
&=& \conv \left( {\cal B}_{1} \circ {\cal B}_{1}\right) \nonumber \\
&=& {\cal O}_{1};
\end{eqnarray}
\label{Condition:JustOneOddWithAlphaLessThanM}
}

\item{if $K_{\alpha}$ is odd for just one value of $\alpha < M$ and $K_M=1$ then
\begin{eqnarray}
\mathcal{S}_{\widetilde{Q}_M}^{EG_N} &=& \conv \left(\left( \mathop{\mathlarger{\mathlarger{\mathlarger{\bigcirc}}}}_{\mathsmaller{\mu < M: K_{\mu} \even}}{\cal T}_{(-1)^{K_{\mu}/2}} \right) \circ {\cal B}_{1} \circ \mathcal{L}_1 \right) \nonumber \\
&=& \conv \left( {\cal B}_{1} \right) \nonumber \\
&=& {\cal B}_{1};
\end{eqnarray}
\label{Condition:JustOneOddWithAlphaLessThanMandKM=1}
}

\item{if $K_{\alpha}$ is odd for just one value of $\alpha < M$ and $K_M>1$ is odd then
\begin{eqnarray}
\mathcal{S}_{\widetilde{Q}_M}^{EG_N} &=& \conv \left(\left( \mathop{\mathlarger{\mathlarger{\mathlarger{\bigcirc}}}}_{\mathsmaller{\mu < M: K_{\mu} \even}}{\cal T}_{(-1)^{K_{\mu}/2}} \right) \circ {\cal B}_{1} \circ \mathcal{T}_\pm \right) \nonumber \\
&=& \conv \left( {\cal B}_{1} \right) \nonumber \\
&=& {\cal B}_{1};
\end{eqnarray}
\label{Condition:JustOneOddWithAlphaLessThanMandKMgeq1Odd}
}

\item{if $M>2$, $K_{\alpha}$ is odd for all values of $\alpha$ and $K_M=1$ then
\begin{eqnarray}
\mathcal{S}_{\widetilde{Q}_M}^{EG_N} &=& \conv \left( \left( \mathop{\mathlarger{\mathlarger{\mathlarger{\bigcirc}}}}_{\mathsmaller{\nu < M: K_{\nu} \odd}}{\cal B}_{1}\right) \circ \mathcal{L}_1 \right) \nonumber \\
&=& \conv \left( \mathop{\mathlarger{\mathlarger{\mathlarger{\bigcirc}}}}_{\mathsmaller{\nu < M: K_{\nu} \odd}}{\cal B}_{1} \right) \nonumber \\
&=& {\cal O}_{1};
\end{eqnarray}
\label{Condition:AllOddKM1Mgeq2}
}

\item{if $M>2$, $K_{\alpha}$ is odd for all values of $\alpha$ and $K_M>1$ then
\begin{eqnarray}
\mathcal{S}_{\widetilde{Q}_M}^{EG_N} &=& \conv \left( \left( \mathop{\mathlarger{\mathlarger{\mathlarger{\bigcirc}}}}_{\mathsmaller{\nu < M: K_{\nu} \odd}}{\cal B}_{1}\right) \circ \mathcal{T}_\pm \right) \nonumber \\
&=& \conv \left( \mathop{\mathlarger{\mathlarger{\mathlarger{\bigcirc}}}}_{\mathsmaller{\nu < M: K_{\nu} \odd}}{\cal B}_{1} \right) \nonumber \\
&=& {\cal O}_{1};
\end{eqnarray}
\label{Condition:AllOddKMgeq1Mgeq2}
}

\item{otherwise,
\begin{eqnarray}
\mathcal{S}_{\widetilde{Q}_M}^{EG_N} &=& \conv \left[ \left( \mathop{\mathlarger{\mathlarger{\mathlarger{\bigcirc}}}}_{\mathsmaller{\mu < M: K_{\mu} \even}}{\cal T}_{(-1)^{K_{\mu}/2}}\right) \circ \left( \mathop{\mathlarger{\mathlarger{\mathlarger{\bigcirc}}}}_{\mathsmaller{\nu < M: K_{\nu} \odd}}{\cal B}_{1}\right) \circ A^{(M)} \right] \nonumber \\
&=& \conv \left[ {\cal T}_{\pm 1} \circ \left( \mathop{\mathlarger{\mathlarger{\mathlarger{\bigcirc}}}}_{\mathsmaller{\nu < M: K_{\nu} \odd}}{\cal B}_{1}\right) \circ A^{(M)} \right] \nonumber \\
%&=& \conv \left[ {\cal T}_{\pm 1}  \circ {\cal B}_{1} \circ \ldots \circ {\cal B}_{1} \right] \nonumber \\
&=& \conv \left( \mathop{\mathlarger{\mathlarger{\mathlarger{\bigcirc}}}}_{\mathsmaller{\nu: K_{\nu} \odd}} {\cal B}_{1} \right) \nonumber \\
&=& {\cal O}_{1}.
\end{eqnarray}
\label{Condition:Otherwise}
}
\end{enumerate}

For any even $N$-qubit system, only a $\widetilde{Q}_M$ partitioning within cases \ref{Condition:AllEven}, \ref{Condition:JustOneOddWithAlphaLessThanMandKM=1}, \ref{Condition:JustOneOddWithAlphaLessThanMandKMgeq1Odd}, \ref{Condition:AllOddKM1Mgeq2}, \ref{Condition:AllOddKMgeq1Mgeq2} and \ref{Condition:Otherwise} may be realised. In cases \ref{Condition:AllEven}, \ref{Condition:JustOneOddWithAlphaLessThanMandKM=1} and \ref{Condition:JustOneOddWithAlphaLessThanMandKMgeq1Odd}, i.e.~when either $K_{\alpha}$ is even for any $\alpha$ or $K_{\alpha}$ is odd for just two values of $\alpha$, one of which is $\alpha=M$, we have that $\mathcal{S}_{\widetilde{Q}_M}^{EG_N}$ is the set ${\cal B}_1$ of all \EGN states. Otherwise, in cases \ref{Condition:AllOddKM1Mgeq2}, \ref{Condition:AllOddKMgeq1Mgeq2} and \ref{Condition:Otherwise}, we have $\mathcal{S}_{\widetilde{Q}_M}^{EG_N} = {\cal O}_{1}$.

For any odd $N$-qubit system, only a $\widetilde{Q}_M$ partitioning within cases \ref{Condition:AllEvenExceptKMequal1}, \ref{Condition:AllEvenExceptKMgeq1Odd}, \ref{Condition:JustOneOddWithAlphaLessThanM}, \ref{Condition:AllOddKM1Mgeq2}, \ref{Condition:AllOddKMgeq1Mgeq2} and \ref{Condition:Otherwise} may be realised. In cases \ref{Condition:AllEvenExceptKMequal1} and \ref{Condition:AllEvenExceptKMgeq1Odd}, i.e.~when $K_{\alpha}$ is odd only for $\alpha=M$, we have that $\mathcal{S}_{\widetilde{Q}_M}^{EG_N}$ is the set ${\cal T}_{(-1)^{(N-1)/2}}$ of all \EGN states. Otherwise, in cases \ref{Condition:JustOneOddWithAlphaLessThanM}, \ref{Condition:AllOddKM1Mgeq2}, \ref{Condition:AllOddKMgeq1Mgeq2} and \ref{Condition:Otherwise}, we have $\mathcal{S}_{\{K_\alpha\}_{\alpha=1}^M}^{\mathcal{M}_{N}^{3}} = {\cal O}_{1}$.

Now we are finally ready to characterise the set of $M$-separable \EGN states $\mathcal{S}_{M}^{EG_N}$ by using Eq.~(\ref{eq:MsepequalconvexhullunionQalphasep}). One can easily see that for any $M\leq \left \lfloor{N/2}\right \rfloor + 1$ one can always find an $M$-partition $\widetilde{Q}_M$ such that $K_{\alpha}$ is odd for at most two values of $\alpha$, one of which is $\alpha=M$, and thus $\mathcal{S}_{\widetilde{Q}_M}^{EG_N} = \mbox{\EGN}$, whereas for any $M> \left \lfloor{N/2}\right \rfloor + 1$ this is impossible and thus $\mathcal{S}_{\widetilde{Q}_M}^{EG_N} ={\cal O}_{1}$ for any $M$-partition $\widetilde{Q}_M$.

\vspace*{-.2cm}
\section*{References}

\vspace*{-.25cm}
\bibliographystyle{iopart-num}

\bibliography{Bibliography}

\end{document}